\documentclass[a4paper,fleqn]{cas-sc}
\usepackage{amsmath,amssymb,bm}
\usepackage{array,booktabs,amsthm}

\usepackage{xcolor}
\usepackage[authoryear,round]{natbib}
\begin{document}
\let\WriteBookmarks\relax
\def\floatpagepagefraction{1}
\def\textpagefraction{.001}
\hypersetup{citecolor=blue,linkcolor=blue,urlcolor=blue}

\shorttitle{A State-Variable Theory of Evolving Material Symmetry}
\shortauthors{Anonymous manuscript}
\RenewDocumentCommand\printorcid{}{}

\title[mode=title]{A State-Variable Theory of Evolving Material Symmetry}

\author[1]{Pengfei Zhu}[orcid=0000-0003-0431-1538]
\cormark[1]
\ead{pengfei.zhu@bam.de}
\credit{Methodology, Investigation, Visualization, Writing -- original draft, Writing -- review \& editing}

\author[2,3]{Leiming Du} 
\credit{Writing -- review \& editing}

\author[3]{Guoqi Zhang}
\credit{Writing -- review \& editing}

\cortext[1]{Corresponding author}

\affiliation[1]{organization={Bundesanstalt f\"ur Materialforschung und -pr\"ufung (BAM)},city={12205 Berlin},country={Germany}}

\affiliation[2]{ organization={Max Planck Institute for Sustainable Materials}, addressline={Max-Planck-Straße 1}, city={Düsseldorf}, postcode={40237}, country={Germany} }

\affiliation[3]{ organization={Department of Microelectronics}, addressline={Delft University of Technology}, city={Mekelweg 4}, postcode={Delft}, postcode={2628 CD}, country={Netherlands}}

\begin{abstract}
Material architecture can evolve during loading, yet its mechanically expressed symmetry depends on both the structural state and the constitutive coupling. We connect these quantities through one parent-invariant free energy with a symmetric traceless internal tensor $\mathbf Q$. Its stabilizer distinguishes isotropic, uniaxial, and biaxial states, and projected gradient kinetics satisfies the structural dissipation inequality. Parent invariance generally guarantees only symmetry inheritance. For an explicit two-coupling elastic potential, we derive a stiffness-contraction reconstruction of $\mathbf Q$ and prove equality of the structural and tangent rotational symmetries for every nonzero coupling pair. The associated acoustic tensor links residual symmetry to protected transverse degeneracies. For a small uniaxial state, the leading squared-speed gap is $|\gamma_2s|\sin^2\vartheta/(2\rho_0)$; coupling selectivity and residual protection can suppress this term. We also derive a directional acoustic-trace reconstruction and state calibrated and nuisance-aware inverse conditions. Dimensionless structural-field, acoustic, transient finite-element, and synthetic inverse examples illustrate the predictions. The framework connects state evolution, mechanically expressed symmetry, and conditional observability without assigning independent dynamics to a group label.
\end{abstract}


\maketitle

\section{Introduction}
\label{sec:introduction}

Material symmetry is one of the central organizing principles of constitutive mechanics~\citep{Zheng1994,Gross1996,Trovalusci1999}. 
It provides the connection between the internal architecture of a material and the admissible form of its constitutive response, constraining both the number and the structure of constitutive coefficients. 
In classical formulations, the material symmetry group is usually prescribed before the constitutive law is constructed and is subsequently treated as fixed~\citep{Brovko2021,Davini2001,Carroll2001,Shai2001}. 
Isotropic, transversely isotropic, orthotropic, and crystalline materials are then represented by distinct symmetry classes, each associated with its own invariant structure and constitutive specialization~\citep{Rice1971,Resende1987,Segev1986}.

This paradigm is natural when the microstructural architecture that supports the constitutive response remains unchanged during the process of interest~\citep{Daymond2007,Yi2006}. 
Many solids, however, do not satisfy this assumption. 
Mechanical loading, temperature variation, chemical environment, phase transformation, damage accumulation, growth, remodeling, or internal rearrangement may alter the preferred directions of the material and, in sufficiently strong cases, change the symmetry class itself~\citep{Stupkiewicz2002,Lu1997,Litewka2003,Margueres2013}. 
Texture may develop from an initially nearly isotropic aggregate, phase transformation may generate or remove crystallographic variants, distributed damage may select preferred orientations, and internal domains may rotate, reorganize, or bifurcate under external driving~\citep{Ambrosi2011,Berlincourt1959,Kreher2002}.

The resulting problem is more general than the evolution of a single material direction. 
A constitutive state may move between qualitatively different symmetry strata. 
For example, a state with no distinguished direction may develop a unique axis, while a uniaxial state may subsequently lose its transverse degeneracy and become biaxial. 
From the viewpoint of constitutive symmetry, these processes correspond schematically to transitions such as
$
G_0
\supset H_1
\supset H_2,
$
where $G_0$ denotes a parent symmetry and the subgroups $H_i$ represent residual symmetries of the evolving structure. 
A theory capable of describing such processes should therefore account not only for the evolution of constitutive parameters within a fixed symmetry class, but also for the evolution of the symmetry constraints themselves.

Several established approaches capture important parts of this problem. 
Structural tensors, fabric tensors, orientation tensors, and related descriptors provide powerful representations of anisotropy and are widely used to encode preferred directions in constitutive laws~\citep{Li2009,Li2015,Jack2007}.
Internal-variable theories allow such descriptors to evolve consistently with thermodynamics, while phase-field and Landau-type formulations provide continuous descriptions of structural transformations through order parameters~\citep{Rosakis2000,Svolos2025,Wu2017,Landau1982}.
In crystal plasticity, damage mechanics, biological remodeling, and transformation mechanics, evolving structural quantities may in turn modify the tangent stiffness and the macroscopic constitutive response~\citep{Arsenlis2002,Acharya2001,Tommasi2009}.
These developments demonstrate that material architecture need not be static.
Yet the evolution of anisotropy and the evolution of material symmetry are not identical concepts. 
A model may continuously rotate a preferred direction while remaining transversely isotropic at every instant, or modify orthotropic moduli without changing the underlying orthotropic symmetry class. 
Conversely, a structural transition may change the isotropy subgroup of the state itself and therefore modify which constitutive transformations remain admissible. 
This distinction suggests that the symmetry class should emerge from the state rather than be imposed independently of it.
A direct strategy would be to switch between separate isotropic, transversely isotropic, orthotropic, or lower-symmetry constitutive laws as the material evolves. 
Such a construction is possible phenomenologically, but it has several conceptual and technical disadvantages. 
It obscures the thermodynamic state variable responsible for the change, introduces artificial boundaries between constitutive descriptions, and makes continuity and differentiability across symmetry transitions difficult to control. 
More importantly, model switching treats the change of symmetry as an external decision rather than as a consequence of the material state.

The alternative developed here is to retain a single parent-symmetric constitutive theory and allow an internal structural state to select the residual symmetry expressed at each material point and time. 
The central idea is therefore not to evolve the material symmetry group directly. 
Instead, one introduces a state variable that transforms covariantly under a fixed parent group and defines the instantaneous material symmetry as the stabilizer of the current state. 
This shifts the role of symmetry from a prescribed label attached to a constitutive equation to an emergent property of a thermodynamic state.
The construction rests on three objects. 
First, a fixed parent group $G_0$ defines the covariance of the constitutive theory. 
Second, a symmetry-bearing internal state $\mathbf Q$ belongs to a representation space of $G_0$. 
Third, the subgroup
$
H_{\mathbf Q}
=
\left\{
g\in G_0:
\rho_{\mathbf Q}(g)\mathbf Q=\mathbf Q
\right\}
$
that leaves the current value of $\mathbf Q$ invariant defines the residual structural symmetry of the state. 
The corresponding orbit
$
\mathcal O_{\mathbf Q}
\simeq
G_0/H_{\mathbf Q}
$
then characterizes the family of symmetry-equivalent states generated by the action of the parent group~\citep{Kobayashi2025,Lami2024,Liu2026,Dao2023}.
This formulation is closely related in mathematical spirit to order-parameter descriptions of spontaneous symmetry breaking, but its role here is constitutive rather than purely classificatory. 
The state $\mathbf Q$ is embedded directly in the free energy, its evolution is constrained by thermodynamics, and its stabilizer determines the symmetry inherited by the constitutive response. 
Hence, the constitutive hierarchy may be expressed as
$
H_{\mathbf Q}
=
\operatorname{Stab}_{G_0}(\mathbf Q),
\Psi=\Psi(\mathbf F,\mathbf Q),
\mathbf S
=
2\partial\Psi/\partial\mathbf C,
\mathbb C
=
4\partial^2\Psi/
(\partial\mathbf C\,\partial\mathbf C).
$
For the minimal three-dimensional realization developed in this work, $\mathbf Q$ is taken to be a symmetric traceless second-order tensor,
$
\mathbf Q\in\operatorname{Sym}_0(3).
$
This space is sufficiently rich to distinguish isotropic, uniaxial, and biaxial structural states while remaining analytically transparent. 
The isotropic state corresponds to $\mathbf Q=\mathbf 0$; a repeated eigenvalue defines a uniaxial state; and three distinct eigenvalues define a biaxial state. 
The associated stabilizers therefore change as the eigenvalue structure of $\mathbf Q$ changes. 
The symmetry classification is consequently encoded in the geometry of the state space rather than introduced as a separate constitutive switch~\citep{Coelho2023,Liu2016,Nissinen2016}.
The use of a symmetric traceless tensor is not intended to imply that every possible material symmetry can be represented by a second-order order parameter. 
Higher-order structural descriptors are required for cubic, tetrahedral, chiral, and other symmetry classes whose defining invariants cannot be captured by $\operatorname{Sym}_0(3)$ alone~\citep{Menzel2003,Karateev2019}.
The present choice should therefore be viewed as a minimal realization designed to expose the mechanics of state-dependent symmetry in a setting that admits explicit constitutive, thermodynamic, and spectral calculations.
A second objective of the theory is to connect symmetry evolution to measurable mechanical observables. 
For a given state $\mathbf Q$, the free energy defines a tangent elasticity tensor $\mathbb C(\mathbf Q)$, from which the acoustic tensor
$
\mathsf A(\widehat{\mathbf k};\mathbf Q)
=
\mathbb C(\mathbf Q)
:
\left(
\widehat{\mathbf k}\otimes\widehat{\mathbf k}
\right)
$
determines incremental elastic-wave speeds and polarizations~\citep{Benallal2004,Moulton2011,Wang2024}.
The symmetry of $\mathbb C$ constrains the spectrum of $\mathsf A$, including the occurrence of symmetry-protected modal degeneracies. 
When the residual symmetry is reduced, previously degenerate branches may split, polarization eigenspaces may separate, and the angular structure of the wave spectrum may change.
This provides a particularly useful bridge between constitutive symmetry and experiment. 
Instead of attempting to observe the internal state $\mathbf Q$ directly, one may interrogate the tangent operator through small-amplitude elastic waves and infer aspects of the evolving symmetry from spectral quantities such as wave-speed splitting, polarization rotation, or the appearance and disappearance of degeneracies~\citep{Liang2025,Liu2018,Goryachev2021}.
In this sense, the wave spectrum acts as a symmetry-sensitive observable of the constitutive state.
The connection between symmetry and spectral degeneracy is classical in wave physics, crystallography, and representation theory~\citep{Csaki2021,Zhang2014,Weissenhofer2025}.
For a fixed propagation direction, the subgroup of material symmetries that leaves the wavevector invariant constrains the acoustic operator and may enforce degeneracies among its eigenmodes. 
The distinction in the present setting is that the relevant material symmetry is itself state dependent. 
The symmetry of the acoustic problem is therefore inherited from the stabilizer of an evolving thermodynamic variable. 
The connection between the evolving structural state and the observable modal response is defined through
$
H_{\mathbf Q}
=
\operatorname{Stab}_{G_0}(\mathbf Q),
\mathbb C=\mathbb C(\mathbf Q),
\mathsf A=\mathsf A(\widehat{\mathbf k};\mathbf Q),
\mathsf A\,\mathbf p_n
=
\rho c_n^2\,\mathbf p_n .
$
Accordingly, changes in \(\mathbf Q\) modify the residual symmetry and tangent elasticity, which in turn determine the acoustic tensor and its modal eigenstructure \(\{c_n,\mathbf p_n\}\).
This perspective also clarifies the role of symmetry breaking in wave splitting. 
A degenerate pair of modes need not split merely because individual constitutive coefficients change; splitting occurs only when the perturbation removes the symmetry responsible for protecting the degeneracy. 
The magnitude and angular dependence of the splitting can therefore carry information about both the amplitude and the symmetry content of the evolving state. 
Near a symmetry-changing transition, this suggests relations of the generic form
$
\Delta c^2
\sim
|\mathbf Q|^p,
$
with the exponent and angular prefactor determined by the constitutive coupling and the symmetry of the active state. 
Such relations provide a mechanically accessible signature of the underlying order parameter without requiring a unique reconstruction of the complete internal structure.
At the same time, spectral inference must be interpreted carefully. 
Wave measurements probe the tangent constitutive operator rather than the structural state in isolation. 
Prestress, conventional elastic anisotropy, damage, density variation, and additional internal variables may produce changes that overlap with those generated by $\mathbf Q$. 
Consequently, the present framework does not assume that the symmetry-bearing state can always be reconstructed uniquely from a limited set of wave measurements. 
Instead, it distinguishes forward symmetry constraints from inverse identifiability and formulates the latter in terms of the sensitivity of measurable spectral quantities to the admissible state and constitutive parameters~\citep{Giljean2024,Lan2018,Castellano2017}.

The theory developed below therefore combines three elements that are often treated separately: a thermodynamic internal state, a group-theoretic definition of residual material symmetry, and a spectral mechanical observable. 
Its central proposition is that material symmetry need not be prescribed as a permanent constitutive label. 
A fixed parent covariance may instead coexist with a time-dependent residual symmetry selected by the current structural state. 
The resulting constitutive law remains a single smooth theory while the isotropy subgroup of the state changes continuously or discontinuously through the state space.
The specific contributions are as follows:
\begin{enumerate}
	\item We assemble a parent-invariant constitutive and kinetic construction in which the stabilizer of a thermodynamic structural state defines residual symmetry. The construction distinguishes orbit reorientation, symmetry-class change, and the symmetry of the complete loaded state.
	\item For an explicit two-coupling elastic potential, we derive a reconstruction identity from two stiffness contractions and prove exact agreement between structural and tangent rotational symmetries. We also state a sufficient positive-energy condition separately from covariance and strong ellipticity.
	\item We connect the evolving state to acoustic little groups and derive the leading uniaxial shear-gap law, $|c_{T1}^2-c_{T2}^2|=|\gamma_2s|\sin^2\vartheta/(2\rho_0)+O(s^2)$. Protected zeros, suppressed first-order couplings, and conditional transition exponents are distinguished explicitly.
\end{enumerate}

\begin{figure}
	\centering
	\includegraphics[width=0.9\textwidth]{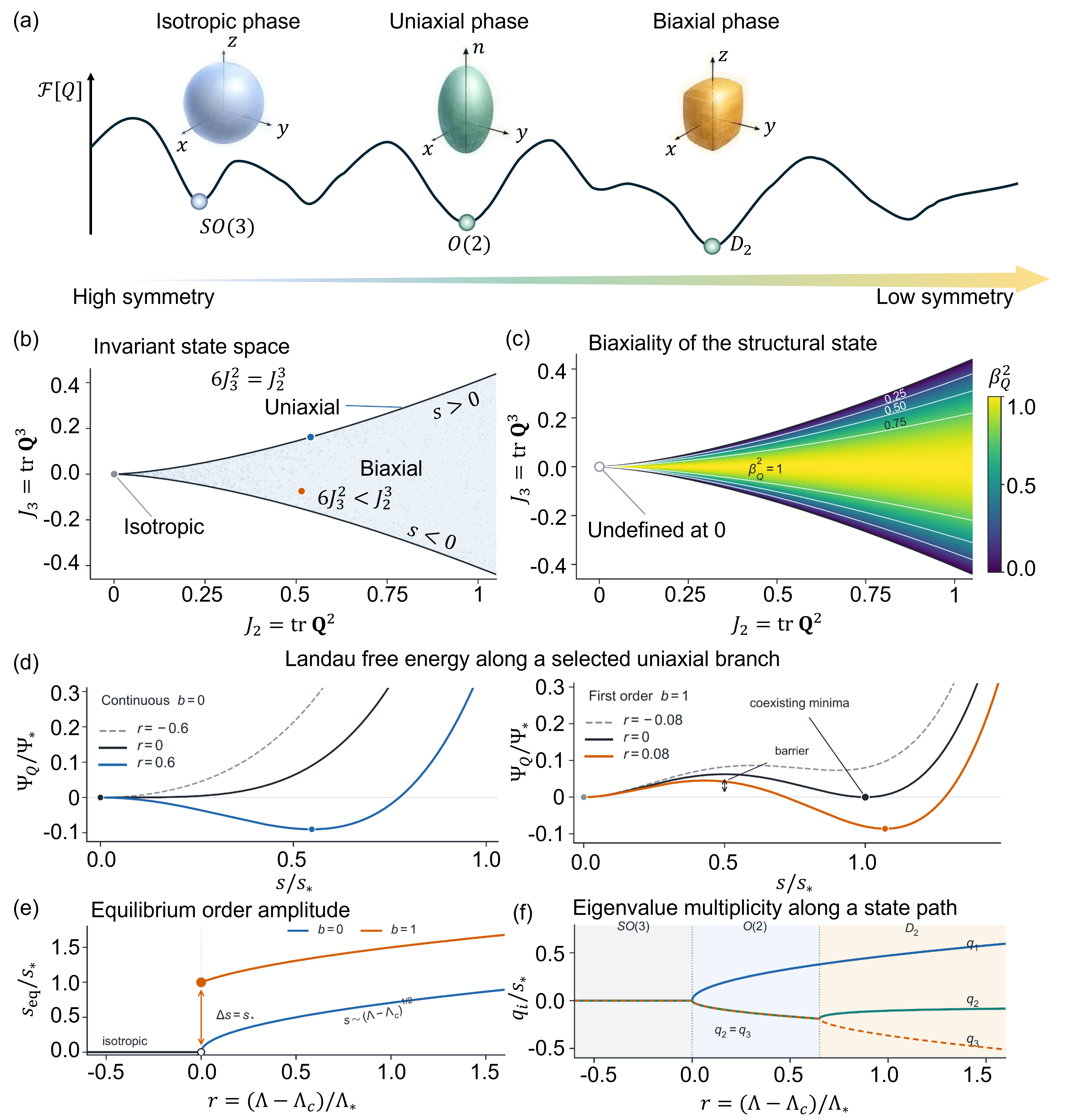}
	\caption{Symmetry stratification and thermodynamic landscape of the structural state.
			(a) Representative isotropic, uniaxial, and biaxial states with residual symmetries $SO(3)$, $O(2)$, and $D_2$.
			(b) Invariant state space $(J_2,J_3)$, where the origin, boundary, and interior correspond to isotropic, uniaxial, and biaxial states, respectively.
			(c) Biaxiality $\beta_Q^2=1-6J_3^2/J_2^3$ over the admissible state space.
			(d) Landau free-energy landscapes illustrating continuous ($b=0$) and first-order ($b=1$) transitions.
			(e) Corresponding equilibrium order amplitude $s_{\rm eq}$, showing continuous onset or a finite jump.
			(f) Eigenvalues of $\mathbf Q$ along a prescribed continuous state path, illustrating the symmetry sequence $SO(3)\rightarrow O(2)\rightarrow D_2$.
			All quantities are dimensionless.}\label{fig1}
\end{figure}

\section{Symmetry-augmented constitutive mechanics}
\label{sec:constitutive_theory}

The formulation separates the choice of a constitutive state from the symmetry expressed by a particular value of that state. The former is fixed once the theory is specified; the latter can change during a process. In what follows, all energy densities are measured per unit reference volume, reference coordinates are Cartesian, and the structural tensor is dimensionless~\citep{Zhu2026_2,Zhu2024}. These conventions make its conjugate force an energy density and the gradient coefficient an energy density multiplied by length squared. The constitutive construction is first given for a general parent-invariant potential and then specialized to an explicit polynomial model. This specialization provides a common basis for the thermodynamic evolution in this section and the wave predictions in Section~\ref{sec:spectral_theory}.

\subsection{State variable, objectivity, and parent-group action}
\label{sec:state_variable}

Let \(\mathcal B_0\) be the reference configuration, \(\mathbf F=\partial\mathbf x/\partial\mathbf X\) the deformation gradient, and \(\mathbf C=\mathbf F^T\mathbf F\) the right Cauchy--Green tensor. We augment a conventional thermomechanical state by a symmetry-bearing internal variable,
\begin{equation}
 \mathcal Z=\{\mathbf C,\theta,\boldsymbol\xi,\mathbf Q,\operatorname{Grad}\mathbf Q\},
 \label{eq:augmented_state}
\end{equation}
where \(\boldsymbol\xi\) denotes other internal variables. The distinction between \(\mathbf Q\) and \(\boldsymbol\xi\) is functional: \(\mathbf Q\) is selected so that its isotropy subgroup carries the structural symmetry relevant to the constitutive response.
For an isotropic parent solid undergoing an isotropic-to-anisotropic transition, the minimal choice is
\begin{equation}
 \mathbf Q\in\operatorname{Sym}_0(3)
 =\{\mathbf A\in\mathbb R^{3\times3}:\mathbf A=\mathbf A^T,\ \operatorname{tr}\mathbf A=0\}.
 \label{eq:Q_space}
\end{equation}
The trace-free condition removes isotropic magnitude from the symmetry descriptor; isotropic volumetric variables may remain in \(\boldsymbol\xi\). A uniaxial state is
\begin{equation}
 \mathbf Q=s\left(\mathbf n\otimes\mathbf n-\frac13\mathbf I\right),
 \qquad \|\mathbf n\|=1,
 \label{eq:uniaxial_Q}
\end{equation}
where \(s\) is the order amplitude and \(\mathbf n\equiv-\mathbf n\) is a director.
The five independent components of \(\mathbf Q\) describe both the strength and the orientation of a second-order anisotropic state. In the uniaxial case, its eigenvalues are \(2s/3,-s/3,-s/3\); changing \(s\) changes the intrinsic ordering, whereas rotating the director changes its orientation in the reference body. A possible microscopic interpretation is a second moment of unoriented material axes,
\begin{equation}
 \mathbf Q=\zeta\left(\langle\mathbf a\otimes\mathbf a\rangle-\frac13\mathbf I\right),
 \qquad |\mathbf a|=1,
 \label{eq:orientation_moment}
\end{equation}
where the averaging is over a representative material volume and \(\zeta\) sets the normalization. This interpretation is useful for textures, aligned networks, or other direction-bearing architectures, but is not imposed on every realization of the theory. If Eq.~\eqref{eq:orientation_moment} is used literally with \(\zeta>0\), its eigenvalues obey \(-\zeta/3\leq q_a\leq2\zeta/3\). The unconstrained polynomial model below is therefore a local phenomenological model; a probability-based realization additionally needs an admissible-state constraint or a potential that enforces those bounds.

Objectivity and material symmetry are distinct~\citep{Zhu2026,Zhu2024_2}. The formulas below use passive material rotations, so $g^T\mathbf Qg$ is a right action; replacing $g$ by $g^{-1}$ gives the conventional left representation. Under a superposed spatial rigid motion, \(\mathbf C\) and the referential tensor \(\mathbf Q\) are unchanged. A material transformation \(g\in G_0\), by contrast, acts on the reference arguments as
\begin{equation}
 \mathbf C\mapsto g^T\mathbf Cg,
 \qquad
 \mathbf Q\mapsto \rho_Q(g)\mathbf Q=g^T\mathbf Qg,
 \qquad
 \boldsymbol\xi\mapsto\rho_\xi(g)\boldsymbol\xi.
 \label{eq:parent_action}
\end{equation}
For \(\mathbf A=\operatorname{Grad}\mathbf Q\), the induced action is
\begin{equation}
 (g\cdot\mathbf A)_{IJK}=g_{PI}g_{QJ}g_{RK}A_{PQR}.
 \label{eq:gradient_action}
\end{equation}

A spatial observer rotation \(\mathbf R_o(t)\) changes \(\mathbf F\) to \(\mathbf R_o\mathbf F\), so that \(\mathbf C\) remains unchanged without rotating the referential structural components. Consequently, an energy written in terms of the state in Eq.~\eqref{eq:augmented_state} is objective. Parent invariance adds a separate restriction: rotating the material arguments together must leave the scalar energy unchanged. The action on \(\operatorname{Grad}\mathbf Q\) rotates all three indices because its first two indices belong to the structural tensor and its third to the material derivative. Omitting the third rotation would make a gradient theory depend on the arbitrary orientation of the reference axes. For a spatially uniform \(g\), this induced action preserves \(|\operatorname{Grad}\mathbf Q|^2\); local rotations \(g(\mathbf X)\) introduce additional derivative terms and are not a symmetry of the present gradient energy.

\begin{figure}
	\centering
	\includegraphics[width=1.0\textwidth]{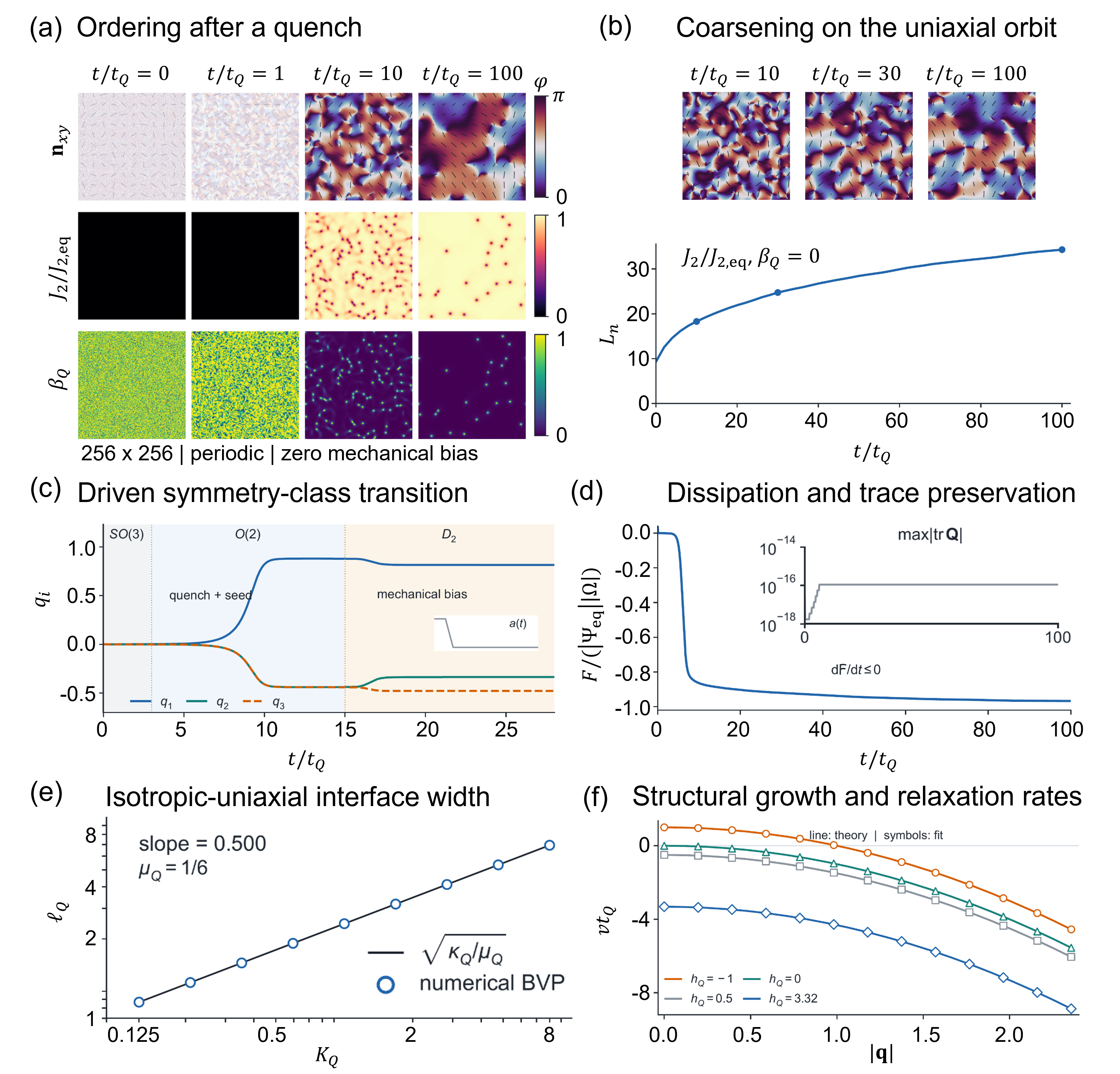}
	\caption{Thermodynamically admissible evolution of material symmetry.
		(a) Ordering of a periodic structural-tensor field following a quench, showing the director field, normalized structural amplitude $J_2/J_{2,\mathrm{eq}}$, and biaxiality $\beta_Q$.
		(b) Coarsening of the orientation texture on a uniaxial orbit at fixed amplitude, characterized by the texture length $L_n$.
		(c) Seeded and driven ordering protocol with isotropic, uniaxial, and biaxial stages, tracked through the eigenvalues $q_i$ of $\mathbf Q$; the interpretation of exact and approximate symmetry is discussed in the text.
		(d) Monotonic decay of the free energy during unforced evolution, with the trace constraint preserved to numerical precision.
		(e) Isotropic--uniaxial interface width $\ell_Q$, confirming the scaling $\ell_Q\sim\sqrt{\kappa_Q/\mu_Q}$.
		(f) Structural growth and relaxation rates as functions of wavenumber, showing agreement between numerical results and $\nu=-\Gamma_Q(h_Q+\kappa_Q|\mathbf q|^2)$.
		All quantities are dimensionless.}\label{fig2}
\end{figure}

\subsection{Residual symmetry and the symmetry-stratified state space}
\label{sec:residual_symmetry}

The residual symmetry of a structural state is its stabilizer in the parent group,
\begin{equation}
 H_{\mathbf Q}=\operatorname{Stab}_{G_0}(\mathbf Q)
 =\{g\in G_0:\rho_Q(g)\mathbf Q=\mathbf Q\}.
 \label{eq:stabilizer}
\end{equation}
If \(\mathbf Q\) is the complete descriptor of a homogeneous bulk structural state, define its residual structural symmetry as \(H_{\mathbf Q}\). Identifying this group with an observable material-response symmetry additionally requires a symmetry-separating constitutive coupling. If several variables carry independent structural information, the complete symmetry is instead
\begin{equation}
 G_{\rm mat}=H_{\mathbf Q}\cap H_{\boldsymbol\xi_s}\cap\cdots.
 \label{eq:intersection_stabilizers}
\end{equation}
For an inhomogeneous gradient theory the structural jet also contributes, so the local complete stabilizer includes $H_{\operatorname{Grad}\mathbf Q}$; the symmetry of an entire specimen must preserve the field and its boundary data. This qualification prevents the symmetry of one descriptor from being mistaken for that of the complete state.
For \(G_0=SO(3)\), write the spectral decomposition
\begin{equation}
 \mathbf Q=\sum_{a=1}^{3}q_a\,\mathbf n_a\otimes\mathbf n_a,
 \qquad \sum_{a=1}^{3}q_a=0.
 \label{eq:Q_spectral}
\end{equation}
The eigenvalue multiplicities give the minimal hierarchy
\begin{equation}
	\mathbf Q=
	\begin{cases}
		\mathbf 0,
		& H_{\mathbf Q}=SO(3),
		\\[1mm]
		\mathbf R\,\mathrm{diag}(\lambda,\lambda,-2\lambda)\,\mathbf R^{T},\quad\lambda\neq0,
		& H_{\mathbf Q}\simeq O(2),
		\\[1mm]
		\mathbf R\,\mathrm{diag}(\lambda_1,\lambda_2,\lambda_3)\,\mathbf R^{T},
		\quad
		\lambda_i\neq\lambda_j\ (i\neq j),\quad\sum_i\lambda_i=0,
		& H_{\mathbf Q}=D_2.
	\end{cases}
	\label{eq:symmetry_hierarchy}
\end{equation}
All groups in Eq.~\eqref{eq:symmetry_hierarchy} are proper rotational stabilizers. The corresponding full orthogonal stabilizers are $O(3)$, $D_{\infty h}$, and $D_{2h}$, since a symmetric second-order tensor cannot distinguish inversion. The embedded \(O(2)\) includes rotations about \(\mathbf n\) and proper rotations that reverse the director; this is why a director has a larger stabilizer than a polar vector.
Here \(\mathbf R\in SO(3)\), and \(D_2\) denotes the identity and the three half-turns about the principal axes, expressed in the eigenframe of \(\mathbf Q\). The subgroup in a rotated frame is its conjugate \(\mathbf R D_2\mathbf R^T\). To see why eigenvalue multiplicities determine these groups, the stabilizer equation can be written as \(\mathbf Qh=h\mathbf Q\). Thus \(h\) must preserve every eigenspace. A repeated pair allows continuous rotations within a two-dimensional eigenspace, whereas three distinct eigenvalues allow only sign changes of the eigenvectors compatible with a proper rotation. At \(\mathbf Q=\mathbf0\), no direction is distinguished. The uniaxial branch requires \(\lambda\neq0\); otherwise it is the isotropic state rather than a state with an axial stabilizer.
Two rotated states \(\mathbf Q_2=g^T\mathbf Q_1g\) have conjugate stabilizers and therefore the same intrinsic symmetry type. The structural state space is consequently partitioned into orbit-type strata,
\begin{equation}
	\operatorname{Sym}_0(3)
	=
	\bigcup_{[H]}
	\left\{
	\mathbf Q\in\operatorname{Sym}_0(3)
	\,\middle|\,
	H_{\mathbf Q}\sim H
	\right\},
	\label{eq:orbit_strata}
\end{equation}
where $H_{\mathbf Q}\sim H$ denotes $H_{\mathbf Q}$ is conjugate to $H$. Motion along a group orbit is reorientation; motion inside one stratum changes amplitude or biaxiality without changing symmetry type; crossing between strata is a material-symmetry transition.
This distinction can also be expressed at the level of rates. If \(\mathbf Q(t)=g(t)^T\mathbf Q_*g(t)\) with fixed \(\mathbf Q_*\), and \(\boldsymbol\Omega=g^T\dot g\) is skew symmetric, then
\begin{equation}
 \dot{\mathbf Q}=\mathbf Q\boldsymbol\Omega-\boldsymbol\Omega\mathbf Q,
 \qquad \dot J_2=0,\qquad\dot J_3=0.
 \label{eq:orbit_rate}
\end{equation}
Reorientation therefore leaves the eigenvalues unchanged. A general dissipative rate has components that change these invariants and need not remain on one orbit~\citep{Zhu2025,Zhu2025_2}. The stabilizer may change discontinuously when eigenvalues coincide even though \(\mathbf Q(t)\), the energy, and the stress vary continuously. This is a change in the classification of the state, not a singular constitutive switch.
The rotational invariants
\begin{equation}
 J_2=\operatorname{tr}(\mathbf Q^2),
 \qquad
 J_3=\operatorname{tr}(\mathbf Q^3)
 \label{eq:Q_invariants}
\end{equation}
parameterize the intrinsic state. For \(\mathbf Q\neq\mathbf0\), the biaxiality measure
\begin{equation}
 \beta_Q^2=1-6\frac{J_3^2}{J_2^3},
 \qquad 0\leq\beta_Q^2\leq1,
 \label{eq:biaxiality}
\end{equation}
vanishes for a uniaxial state and is positive for a generic biaxial state. The isotropic origin is treated separately.

The sufficiency of \(J_2\) and \(J_3\) follows from the characteristic identity \(\mathbf Q^3-(J_2/2)\mathbf Q-(J_3/3)\mathbf I=\mathbf0\). These invariants determine the unordered eigenvalues and hence the orbit under \(SO(3)\), although they do not determine the orientation of its eigenframe. Their admissible domain satisfies \(6J_3^2\leq J_2^3\), with the uniaxial states on its boundary. The quantity \(\beta_Q^2\) characterizes biaxiality in this invariant domain; it is not a Euclidean distance in tensor space. In particular, the limit of \(\beta_Q^2\) at the origin depends on the approach path, so isotropy is identified by \(J_2=0\) rather than by assigning a limiting biaxiality there. This separation of orientation, amplitude, and symmetry type is essential when comparing structural measurements with directional mechanical response.

Figure~\ref{fig1}(a--c) visualizes the invariant domain: the origin, repeated-eigenvalue boundary, and interior represent isotropic, uniaxial, and biaxial states. The ellipsoids depict the spectrum of $\exp(1.4\mathbf Q)$; they are structural glyphs rather than specimen deformations. Panel (f) is a prescribed continuous tensor path crossing these strata, not an equilibrium biaxial branch of the unloaded quartic potential. Its parameterization is given in Supplementary Section~\ref{sec:geometry}.

\begin{figure}
	\centering
	\includegraphics[width=1.0\textwidth]{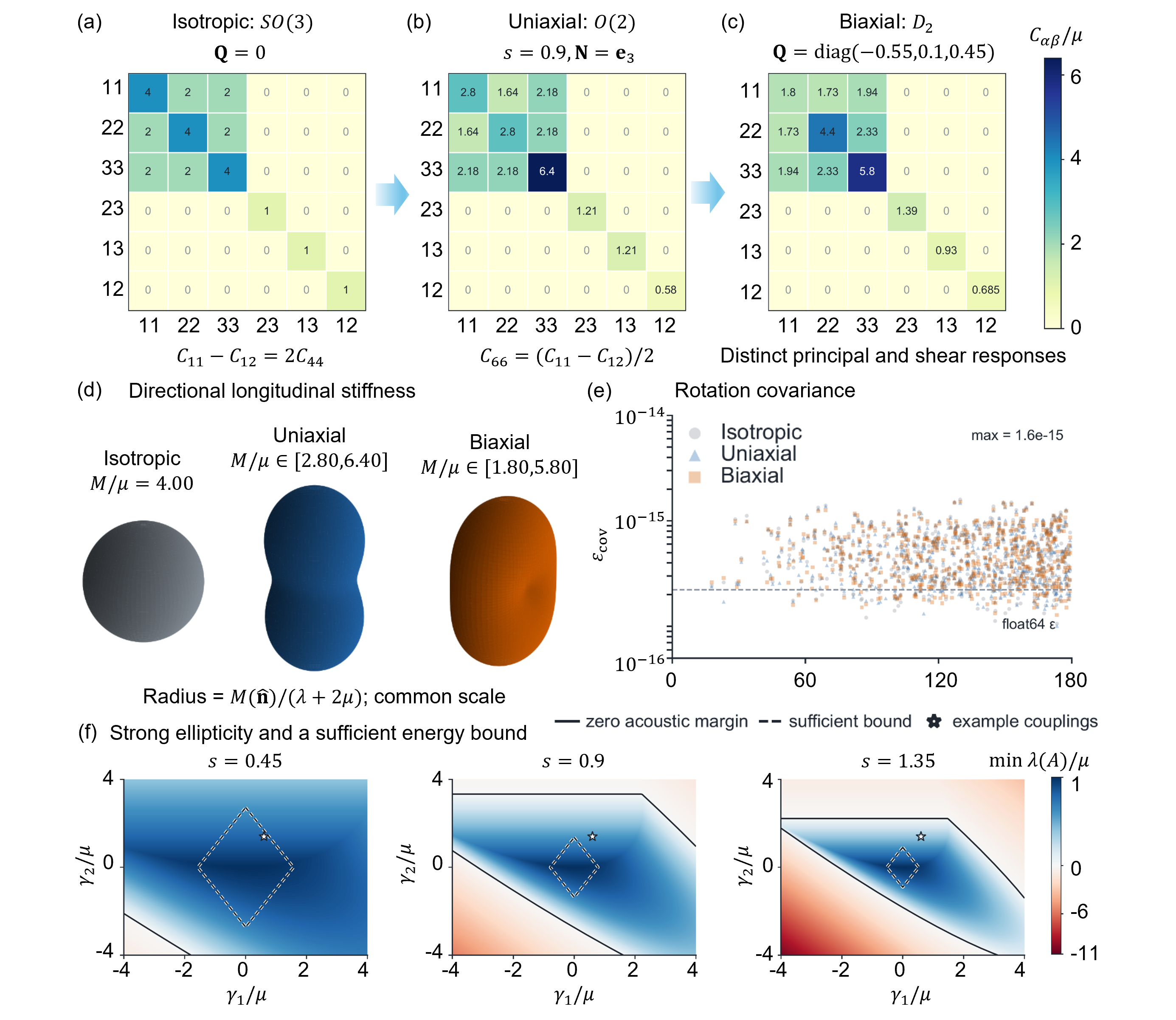}
	\caption{Inheritance of structural symmetry by the constitutive tangent.
		(a--c) Engineering Voigt stiffness matrices for isotropic, uniaxial, and biaxial structural states, illustrating the symmetry-dependent constitutive response.
		(d) Directional longitudinal stiffness $M(\hat{\mathbf n})=(\hat{\mathbf n}\otimes\hat{\mathbf n}):\mathbb C:(\hat{\mathbf n}\otimes\hat{\mathbf n})$, showing the transition from isotropic to anisotropic response.
		(e) Rotational covariance errors for randomly rotated structural states, confirming the parent-group invariance of the constitutive formulation to machine precision.
		(f) Strong-ellipticity domains in the $(\gamma_1,\gamma_2)$ parameter space for increasing structural amplitudes, comparing the acoustic stability boundary with a sufficient positive-energy bound.
		All quantities are dimensionless.}\label{fig3}
\end{figure}

\subsection{One parent-invariant potential and state-dependent response}
\label{sec:master_potential}

The constitutive law is not changed when \(H_{\mathbf Q}\) changes. Instead, one master free-energy density is invariant under the fixed parent group:
\begin{equation}
 \Psi(g^T\mathbf Cg,\theta,\rho_\xi(g)\boldsymbol\xi,
 g^T\mathbf Qg,g\cdot\operatorname{Grad}\mathbf Q)
 =\Psi(\mathbf C,\theta,\boldsymbol\xi,\mathbf Q,\operatorname{Grad}\mathbf Q)
 \quad \forall g\in G_0.
 \label{eq:parent_invariance}
\end{equation}
For an isotropic parent, useful mixed invariants include
\begin{equation}
 K_1=\operatorname{tr}(\mathbf C\mathbf Q),\qquad
 K_2=\operatorname{tr}(\mathbf C\mathbf Q^2),\qquad
 K_3=\operatorname{tr}(\mathbf C^2\mathbf Q).
 \label{eq:mixed_invariants}
\end{equation}
These invariants illustrate how the mechanical and structural arguments can be coupled without selecting preferred reference axes in advance. They are not asserted to form a complete integrity basis for an arbitrary joint function of two symmetric tensors; further mixed invariants, such as \(\operatorname{tr}(\mathbf C^2\mathbf Q^2)\), may be needed. A scalar function built from a suitable collection generates a continuous family of state-dependent responses. Parent invariance restricts the allowed tensor form, while the scalar coefficients remain material-dependent functions of temperature and other invariant state variables.
The second Piola--Kirchhoff stress and material tangent are
\begin{equation}
 \mathbf S=2\frac{\partial\Psi}{\partial\mathbf C},
 \qquad
 \mathbb C=4\frac{\partial^2\Psi}{\partial\mathbf C\,\partial\mathbf C}.
 \label{eq:stress_tangent}
\end{equation}
Differentiating Eq.~\eqref{eq:parent_invariance} shows that these tensor-valued responses are equivariant. For the frozen-state constitutive function over all strains, every \(h\in H_{\mathbf Q}\) leaves the structural argument unchanged and therefore constrains the response. Parent invariance guarantees
\begin{equation}
 H_{\mathbf Q}\subseteq G_{\rm const}(\mathbf Q),
 \label{eq:symmetry_inheritance}
\end{equation}
where \(G_{\rm const}\) is the proper rotational symmetry group of that frozen-state constitutive function. A tangent evaluated at an arbitrary deformed state inherits only the intersection with the deformation stabilizer, as specified in Eq.~\eqref{eq:base_state_symmetry}. Equality requires symmetry-separating coupling: if \(\Psi\) does not couple mechanics to \(\mathbf Q\), the mechanical response may retain a larger accidental symmetry.
The inheritance statement applies when the remaining structural arguments are also preserved, or when only \(\mathbf Q\) enters the bulk mechanical energy. With independent anisotropic variables, the relevant subgroup is the intersection in Eq.~\eqref{eq:intersection_stabilizers}. For example, setting the mixed coefficients to zero leaves a mechanically isotropic solid even at a nonzero \(\mathbf Q\). A response depending only on \(\mathbf Q^2\) can likewise conceal distinctions present in \(\mathbf Q\). Structural symmetry and mechanically observable symmetry should therefore be compared through the actual coupling, rather than identified solely from the choice of state space.
At small strain \(\mathbf E=(\mathbf C-\mathbf I)/2\), a compact parent-isotropic realization is
\begin{align}
 \Psi={}&\frac{\lambda}{2}(\operatorname{tr}\mathbf E)^2+\mu\,\mathbf E:\mathbf E
 +\gamma_1(\operatorname{tr}\mathbf E)(\mathbf E:\mathbf Q)
 +\gamma_2\operatorname{tr}(\mathbf Q\mathbf E^2)
 \nonumber\\
 &+\Psi_Q(\mathbf Q)+\frac{\kappa_Q}{2}|\operatorname{Grad}\mathbf Q|^2.
 \label{eq:minimal_energy}
\end{align}
The two mixed terms vanish to first order in \(\mathbf E\), so \(\mathbf E=\mathbf0\) remains stress free, but their second derivatives make the tangent stiffness depend on \(\mathbf Q\). If ordering produces a transformation strain, the mechanical energy can instead be written using \(\mathbf E-\mathbf E^\star(\mathbf Q)\), with
\begin{equation}
 \mathbf E^\star(\mathbf Q)=\alpha_1\mathbf Q+\alpha_2\operatorname{dev}(\mathbf Q^2)+\cdots.
 \label{eq:transformation_strain}
\end{equation}
These are different physical assumptions: Eq.~\eqref{eq:minimal_energy} describes symmetry-dependent stiffness about a fixed natural state, whereas Eq.~\eqref{eq:transformation_strain} permits an evolving natural configuration.
For the fixed-natural-state realization, the constitutive derivatives can be evaluated explicitly. With \(\theta\), \(\boldsymbol\xi\), and \(\mathbf Q\) held fixed, differentiation with respect to \(\mathbf E\) gives the same second Piola--Kirchhoff stress as Eq.~\eqref{eq:stress_tangent},
\begin{align}
 \mathbf S={}&\lambda(\operatorname{tr}\mathbf E)\mathbf I+2\mu\mathbf E
 +\gamma_1\big[(\mathbf E:\mathbf Q)\mathbf I+(\operatorname{tr}\mathbf E)\mathbf Q\big]
 +\gamma_2(\mathbf Q\mathbf E+\mathbf E\mathbf Q),
 \label{eq:explicit_stress}\\
 \mathbb C:\mathbf H={}&\lambda(\operatorname{tr}\mathbf H)\mathbf I+2\mu\mathbf H
 +\gamma_1\big[(\mathbf H:\mathbf Q)\mathbf I+(\operatorname{tr}\mathbf H)\mathbf Q\big]
 +\gamma_2(\mathbf Q\mathbf H+\mathbf H\mathbf Q),
 \label{eq:explicit_tangent}
\end{align}
for any symmetric strain increment \(\mathbf H\). The stress is zero at \(\mathbf E=\mathbf0\) for every fixed structural state, but the incremental response at that point depends linearly on \(\mathbf Q\). In components, the structural contribution to the tangent is
\begin{equation}
 \delta C_{ijkl}=\gamma_1(\delta_{ij}Q_{kl}+Q_{ij}\delta_{kl})
 +\frac{\gamma_2}{2}(Q_{ik}\delta_{jl}+Q_{il}\delta_{jk}
 +\delta_{ik}Q_{jl}+\delta_{il}Q_{jk}).
 \label{eq:explicit_tangent_components}
\end{equation}
Its minor and major symmetries follow from the existence of the scalar potential. These relations also show why two different couplings are retained: the \(\gamma_1\) term couples volumetric and directional strain, whereas \(\gamma_2\) changes the response to shear strains. Their distinct effects on elastic waves will be identifiable after projection onto the transverse eigenspace.

For the tangent in Eq.~\eqref{eq:explicit_tangent_components}, including its isotropic part, let $(\gamma_1,\gamma_2)\neq(0,0)$. Then $\mathbf Q\mapsto\mathbb C(\mathbf Q)$ is injective for fixed coefficients and
\begin{equation}
 \operatorname{Stab}_{SO(3)}\mathbb C(\mathbf Q)=H_{\mathbf Q}.
 \label{eq:exact_tangent_symmetry}
\end{equation}

Introduce the dilatation and Voigt contractions $D_{ij}=C_{ijkk}$ and $V_{ik}=C_{ijkj}$~\citep{Cowin1987}. Direct contraction gives
\begin{equation}
 \operatorname{dev}\mathbf D=a_D\mathbf Q,\qquad a_D=3\gamma_1+2\gamma_2,\nonumber\\
 \operatorname{dev}\mathbf V=a_V\mathbf Q,\qquad a_V=2\gamma_1+\tfrac52\gamma_2.
 \label{eq:stiffness_contractions}
\end{equation}
The coefficient matrix mapping $(\gamma_1,\gamma_2)$ to $(a_D,a_V)$ has determinant $7/2$, so both contractions vanish identically only when both couplings vanish. Consequently,
\begin{equation}
 \mathbf Q=\frac{a_D\operatorname{dev}\mathbf D+a_V\operatorname{dev}\mathbf V}
 {a_D^2+a_V^2}.
 \label{eq:Q_from_stiffness}
\end{equation}
Contractions and the deviatoric projection commute with orthogonal transformations. Every rotation preserving $\mathbb C$ therefore preserves the reconstructed $\mathbf Q$, while parent covariance gives the reverse inclusion. This proves Eq.~\eqref{eq:exact_tangent_symmetry}.

This result is specific to the calibrated two-coupling potential. It demonstrates that a nonzero coupling transmits the structural symmetry exactly to the complete tangent even when a particular wave gap is blind to one coefficient. It does not imply that a limited collection of speeds determines every component of that tangent. The reconstruction is also poorly conditioned as both couplings approach zero~\citep{Zhu2026a,Zhu2026b}.

Parent invariance alone does not imply mechanical stability. In the isotropic limit, positive strain energy requires \(\mu>0\) and \(3\lambda+2\mu>0\). Writing \(m_*=\min\{2\mu,3\lambda+2\mu\}\), a sufficient condition for positive definiteness of the tangent of this polynomial model is
\begin{equation}
 2(\sqrt3|\gamma_1|+|\gamma_2|)\|\mathbf Q\|<m_*.
 \label{eq:sufficient_elastic_stability}
\end{equation}
Indeed, \(|\operatorname{tr}\mathbf H|\leq\sqrt3\|\mathbf H\|\), \(|\mathbf H:\mathbf Q|\leq\|\mathbf H\|\|\mathbf Q\|\), and \(|\operatorname{tr}(\mathbf Q\mathbf H^2)|\leq\|\mathbf Q\|\|\mathbf H\|^2\) bound the perturbation of \(\mathbf H:\mathbb C:\mathbf H\). This bound is sufficient rather than necessary, but makes explicit the neighborhood in which the minimal linear-in-\(\mathbf Q\) stiffness model is guaranteed to remain stable. Larger structural amplitudes may require higher-order invariant couplings calibrated to retain stability.

Figure~\ref{fig3} evaluates the same potential at $\lambda=2$, $\mu=1$, $\gamma_1=0.6$, and $\gamma_2=1.4$, for $\mathbf Q=\mathbf0$, $\operatorname{diag}(-0.3,-0.3,0.6)$, and $\operatorname{diag}(-0.55,0.1,0.45)$. The matrices exhibit isotropic, transversely isotropic, and orthotropic structure, respectively. The latter remains a restricted family within the nine-modulus orthotropic class. The directional stiffness in panel (d) is $(\mathbf m\otimes\mathbf m):\mathbb C:(\mathbf m\otimes\mathbf m)$, which need not equal a quasi-longitudinal acoustic eigenvalue. The reported covariance residual is approximately $1.6\times10^{-15}$. Panel (f) distinguishes strong ellipticity from the sufficient positive-energy bound; angular sampling of acoustic eigenvalues is a numerical diagnostic rather than an analytical certificate over the entire sphere. Definitions and additional checks are given in Supplementary Section~\ref{sec:stiffness}.

\begin{figure}
	\centering
	\includegraphics[width=1.0\textwidth]{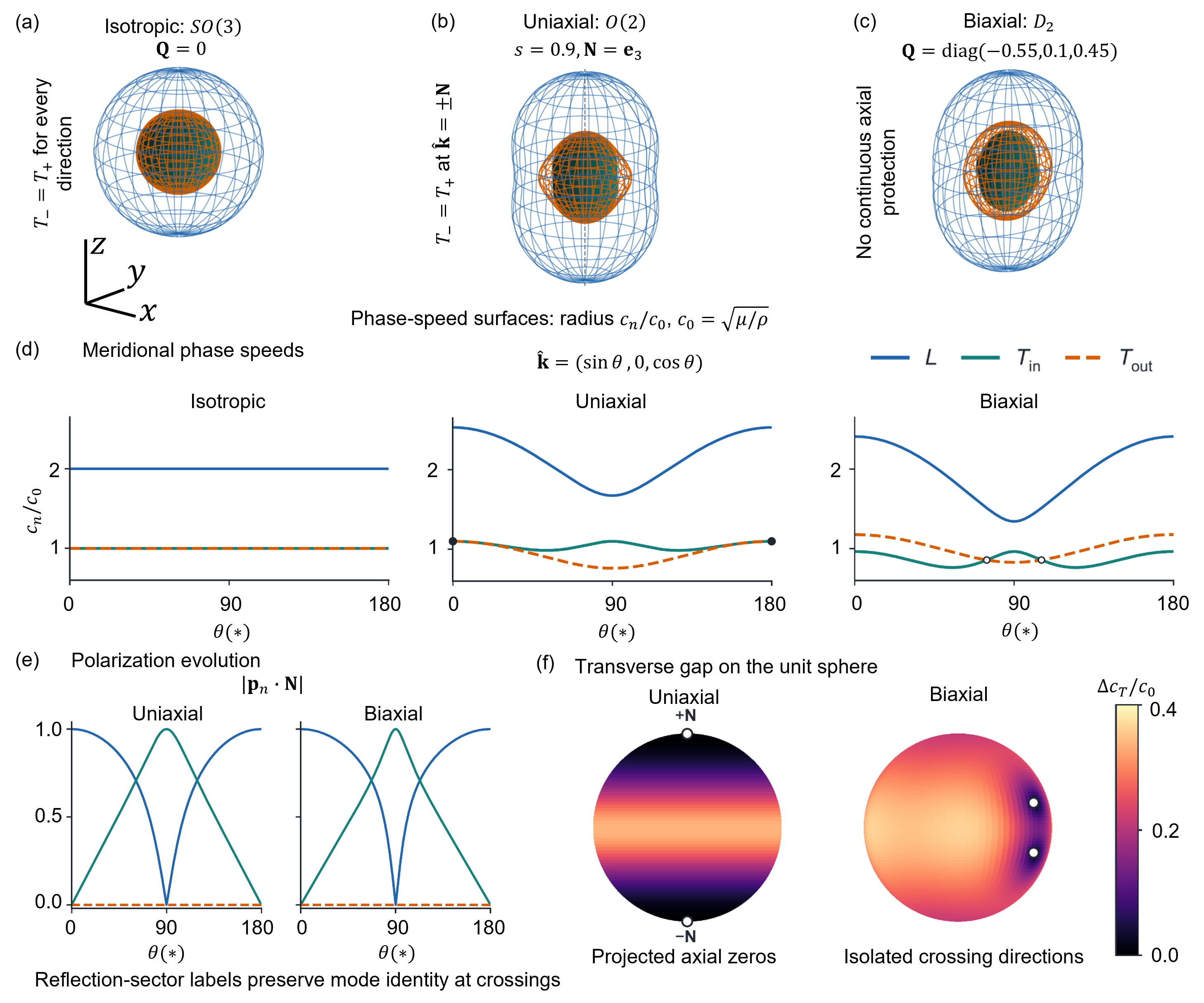}
	\caption{Symmetry-dependent acoustic spectra and modal degeneracies.
		(a--c) Three-dimensional phase-speed surfaces for isotropic, uniaxial, and biaxial structural states, illustrating the progressive lifting of transverse-mode degeneracy.
		(d) Directional longitudinal and transverse phase speeds, showing symmetry-protected degeneracies and isolated mode crossings.
		(e) Evolution of modal polarizations along the meridional propagation plane, characterized by their projections onto the structural axis.
		(f) Angular distributions of the transverse phase-speed gap $\Delta c_T/c_0$, revealing protected axial degeneracies in uniaxial states and isolated crossing directions in biaxial states.
		Phase speeds are normalized by $c_0=\sqrt{\mu/\rho}$, and all quantities are dimensionless.}\label{fig4}
\end{figure}

\subsection{Thermodynamic force and evolution law}
\label{sec:thermodynamic_evolution}

The evolution law is obtained from the same energy that generates the mechanical response. To make the constraint explicit, introduce the referential entropy density \(\eta\) and heat flux \(\mathbf h_0\). The free-energy imbalance includes the energetic structural microtraction because the state contains \(\operatorname{Grad}\mathbf Q\). After integration by parts over an arbitrary material part and localization, its bulk terms are
\begin{equation}
 \mathcal D=\left(\frac12\mathbf S-\Psi_{,\mathbf C}\right):\dot{\mathbf C}
 -(\eta+\Psi_{,\theta})\dot\theta
 -\Psi_{,\boldsymbol\xi}\mathbin{\cdot}\dot{\boldsymbol\xi}
 +\left(-\Psi_{,\mathbf Q}+\operatorname{Div}\mathbf M_Q\right):\dot{\mathbf Q}
 -\frac{\mathbf h_0\cdot\operatorname{Grad}\theta}{\theta}\geq0.
 \label{eq:full_reduced_imbalance}
\end{equation}
Independent reversible strain and temperature rates give Eq.~\eqref{eq:stress_tangent} and \(\eta=-\Psi_{,\theta}\). The remaining terms are dissipative contributions from the internal variables, the structural field, and heat conduction. The present structural closure assumes that the other dissipative mechanisms are assigned their own admissible laws; it does not absorb plasticity or damage dissipation into \(\mathbf Q\). This distinction permits additional mechanisms to be introduced without changing the interpretation of the symmetry-bearing state.
Let
\begin{equation}
 \mathcal F[\mathbf Q]=\int_{\mathcal B_0}\Psi(\mathbf C,\theta,\boldsymbol\xi,
 \mathbf Q,\operatorname{Grad}\mathbf Q)\,\mathrm dV.
 \label{eq:free_energy_functional}
\end{equation}
The structural microstress and bulk driving force are
\begin{equation}
 \mathbf M_Q=\frac{\partial\Psi}{\partial(\operatorname{Grad}\mathbf Q)},
 \qquad
 \mathbf A_Q=-\frac{\partial\Psi}{\partial\mathbf Q}+\operatorname{Div}\mathbf M_Q
 =-\frac{\delta\mathcal F}{\delta\mathbf Q}.
 \label{eq:structural_force}
\end{equation}
Because admissible rates belong to \(\operatorname{Sym}_0(3)\), only the projected force
\begin{equation}
 \mathbf A_Q^0=\mathbb P_0[\mathbf A_Q],
 \qquad
 \mathbb P_0[\mathbf A]=\operatorname{sym}\mathbf A-\frac{1}{3}\operatorname{tr}(\mathbf A)\mathbf I
 \label{eq:traceless_projection}
\end{equation}
can perform structural work. A dissipative closure is
\begin{equation}
 \dot{\mathbf Q}=\mathbb M_Q:\mathbf A_Q^0,
 \qquad
 \mathbf A:\mathbb M_Q:\mathbf A\geq0
 \quad\forall\mathbf A\in\operatorname{Sym}_0(3),
 \label{eq:kinetic_law}
\end{equation}
where $\mathbb M_Q$ maps $\operatorname{Sym}_0(3)$ into itself, which gives \(\mathcal D_Q=\mathbf A_Q^0:\dot{\mathbf Q}\geq0\). For isotropic mobility \(\mathbb M_Q=\Gamma_Q\mathbb I_0\),
\begin{equation}
 \dot{\mathbf Q}=-\Gamma_Q\mathbb P_0\left[\frac{\delta\mathcal F}{\delta\mathbf Q}\right],
 \qquad \Gamma_Q\geq0.
 \label{eq:allen_cahn_Q}
\end{equation}
This equation evolves the structural state; the symmetry changes only as the derived stabilizer \(H_{\mathbf Q(t)}\) changes.
The projection is an orthogonal projection in the Frobenius inner product. Consequently, any spherical part of \(\mathbf A_Q\) performs no work on an admissible rate, and the trace constraint is preserved if it holds initially. The mobility must additionally respect parent covariance: for any admissible \(\mathbf A\),
\begin{equation}
 \mathbb M_Q(g^T\mathbf Qg): (g^T\mathbf Ag)
 =g^T(\mathbb M_Q(\mathbf Q):\mathbf A)g,
 \label{eq:mobility_covariance}
\end{equation}
with all other state arguments transformed consistently when present. Isotropic scalar mobility meets this requirement automatically. Under fixed mechanical and thermal arguments and zero structural boundary power, the evolution satisfies
\begin{equation}
 \frac{\mathrm d\mathcal F}{\mathrm dt}
 =-\int_{\mathcal B_0}\mathbf A_Q^0:\mathbb M_Q:\mathbf A_Q^0\,\mathrm dV\leq0.
 \label{eq:energy_decay}
\end{equation}
External loading or a changing temperature supplies additional power, so monotonic decay of the structural energy is not claimed for a driven process. Natural boundary conditions are \(\mathbf M_Q\mathbf N=\mathbf0\), while prescribed structural orientation or amplitude can be imposed by a Dirichlet condition. The boundary choice is part of the physical problem because surface anchoring can select a texture even for an isotropic bulk parent.
For the minimal Landau potential
\begin{equation}
 \Psi_Q=\frac{a}{2}J_2-\frac{b}{3}J_3+\frac{c}{4}J_2^2,
 \qquad c>0,
 \label{eq:Landau_potential}
\end{equation}
and the isotropic gradient term in Eq.~\eqref{eq:minimal_energy}, the evolution becomes
\begin{equation}
 \dot{\mathbf Q}=\Gamma_Q\left[-a\mathbf Q+b\operatorname{dev}(\mathbf Q^2)
 -c(\mathbf Q:\mathbf Q)\mathbf Q+\kappa_Q\Delta\mathbf Q+\mathbf A_{\rm mech}^0\right],
 \label{eq:explicit_Q_evolution}
\end{equation}
where \(\mathbf A_{\rm mech}^0\) is generated by the mixed mechanical terms.
For Eq.~\eqref{eq:minimal_energy}, this mechanical contribution is explicitly
\begin{equation}
 \mathbf A_{\rm mech}^0=-\gamma_1(\operatorname{tr}\mathbf E)\operatorname{dev}\mathbf E
 -\gamma_2\operatorname{dev}(\mathbf E^2).
 \label{eq:mechanical_structural_force}
\end{equation}
Loading can therefore bias both the amplitude and the eigenframe of the structural state. At zero strain this bias vanishes, whereas a deviatoric strain generally supplies a direction-bearing force. An imposed strain acting as a conjugate field can round an otherwise sharp zero-field transition; the critical laws derived below are consequently stated for an unbiased homogeneous state. In particular, a symmetry transition inferred from a loaded specimen cannot be interpreted without distinguishing intrinsic ordering from the symmetry imposed by the deformation.

Figure~\ref{fig2}(a,d) illustrates unforced ordering with $(a,b,c,\kappa_Q,\Gamma_Q)=(-1,3,2,1,1)$ on a periodic domain. The energy decreases and the reported trace residual remains of order $10^{-16}$. Panel (b) uses a separate mobility constrained to a uniaxial orbit: texture coarsening then changes orientation without changing the local symmetry type. Panel (c) uses a seed and a subsequent mechanical bias to generate an ordering and biaxiality protocol. These imposed interventions must be distinguished from the unforced gradient flow. In a spatially varying field, $H_{\mathbf Q(\mathbf X)}$ is a local descriptor; the gradient, boundary conditions, and texture need not share that stabilizer. The numerical protocols are described in Supplementary Section~\ref{sec:kinetics}.

An exactly symmetric deterministic state does not spontaneously leave its invariant subspace. For a homogeneous equivariant rate $\dot{\mathbf Q}=\mathcal V(\mathbf Q,t)$ with unique solutions and parent-invariant controls, $h^T\mathbf Q(0)h=\mathbf Q(0)$ implies $h^T\mathbf Q(t)h=\mathbf Q(t)$ by equivariance and uniqueness. In particular, the unbiased zero state remains zero even after it becomes linearly unstable. Ordering from that state requires a perturbation, noise, or a symmetry-breaking field. Likewise, eigenvalues that approach equality numerically should not be called exactly equal without a tolerance and convergence check. Figure~\ref{fig2}(c) is therefore an externally seeded and driven protocol, rather than a demonstration of deterministic symmetry loss from exact isotropy.

 A characteristic diffuse-interface width follows from the balance of local curvature \(\mu_Q\) and gradient energy,
\begin{equation}
 \ell_Q\sim\sqrt{\kappa_Q/\mu_Q}.
 \label{eq:interface_length}
\end{equation}

The associated time scale is equally important. Linearizing about a homogeneous stationary state gives a projected structural Hessian \(\mathbb H_Q\). For a Fourier perturbation \(\delta\mathbf Q=\mathbf V\exp(i\mathbf q\cdot\mathbf X+\nu t)\) in an eigendirection with curvature \(h_Q\), scalar mobility gives
\begin{equation}
 \nu=-\Gamma_Q(h_Q+\kappa_Q|\mathbf q|^2),
 \qquad \tau_Q(\mathbf q)=\big[\Gamma_Q(h_Q+\kappa_Q|\mathbf q|^2)\big]^{-1}
 \label{eq:structural_linear_relaxation}
\end{equation}
when the denominator is positive. Positive curvature causes relaxation, negative curvature causes long-wavelength growth, and the gradient term suppresses sufficiently short wavelengths. The estimate for \(\ell_Q\) uses a positive restoring curvature in an amplitude or interface-relevant sector. Pure orientation variations along an orbit of a homogeneous parent-invariant potential have zero local restoring curvature; a finite orientational length then requires gradients, boundary anchoring, or an external bias. Thus an interface width and an orientation-relaxation length should not be assigned the same curvature without checking the mode being perturbed.
The coexistence-interface and linear-rate tests in Fig.~\ref{fig2}(e,f) support these length and time scales. The amplitude interface has a logistic profile at $a_t=b^2/(27c)$ and $s_t=b/(3c)$, with $\ell_Q=\sqrt{\kappa_Q/a_t}$. The orientational sector instead has zero homogeneous curvature in the absence of anchoring or bias. The reduction and numerical width definition are given in Supplementary Section~\ref{sec:interfaces}.

Along the uniaxial ansatz in Eq.~\eqref{eq:uniaxial_Q},
\begin{equation}
 J_2=\frac{2}{3}s^2,
 \qquad J_3=\frac{2}{9}s^3,
 \qquad
 \Psi_Q(s)=\frac{a}{3}s^2-\frac{2b}{27}s^3+\frac{c}{9}s^4.
 \label{eq:uniaxial_potential}
\end{equation}
The homogeneous stationary branches satisfy
\begin{equation}
 s=0
 \quad\text{or}\quad
 2cs^2-bs+3a=0.
 \label{eq:uniaxial_branches}
\end{equation}
The cubic invariant permits a first-order transition, whereas its absence produces the familiar continuous-transition limit. These branches illustrate, rather than universally prescribe, the material energetics.
More explicitly, the nonzero uniaxial branches are \(s_\pm=[b\pm\sqrt{b^2-24ac}]/(4c)\). For \(b>0\), comparison with the isotropic energy gives coexistence at \(a_t=b^2/(27c)\) and an amplitude jump \(s_t=b/(3c)\); the ordered stationary branches first appear at \(a=b^2/(24c)\), while the isotropic state loses local stability at \(a=0\). These different thresholds distinguish equilibrium coexistence from metastability and spinodal loss of stability. When \(b=0\), a chosen uniaxial branch instead satisfies
\begin{equation}
 a(\Lambda)=a_0(\Lambda_c-\Lambda),\qquad
 |s|=\left[\frac{3a_0}{2c}(\Lambda-\Lambda_c)\right]^{1/2},
 \quad a_0>0,\quad\Lambda>\Lambda_c.
 \label{eq:continuous_order_scaling}
\end{equation}
Here \(\Lambda\) is an invariant control parameter and the result assumes no mechanical conjugate field. The exponent follows from this quartic mean-field potential, not from symmetry alone. Moreover, at \(b=0\) the local potential depends only on \(J_2\), so it does not select uniaxiality over biaxiality: all tensors with \(J_2=-a/c\) are degenerate minima. Equation~\eqref{eq:continuous_order_scaling} describes the selected uniaxial path, which needs an additional selection mechanism if used as a material prediction. This qualification is important when connecting an amplitude law to a particular spectral angular pattern.
Figures~\ref{fig1}(d,e) compare continuous onset on the selected $b=0$ uniaxial path with first-order coexistence for $b>0$. At coexistence the ordered and isotropic minima have equal energy; the jump occurs before isotropic spinodal instability. The normalizations and plotted scalar sections are given in Supplementary Section~\ref{sec:landau}.

The complete constitutive architecture is therefore
\begin{equation}
	\bigl(G_0,\mathcal V_Q,\rho_Q,\Psi\bigr)
	\mapsto
	\bigl\{\mathbf Q(\mathbf X,t),\,H_{\mathbf Q},\,\mathbf S,\,\mathbb C\bigr\}.
	\label{eq:constitutive_chain}
\end{equation}

The mapping summarizes a constitutive construction rather than a unique solution of an evolution problem. A trajectory \(\mathbf Q(\mathbf X,t)\) also requires the kinetic law, initial and boundary data, and mechanical and thermal loading. Once those are supplied, the stabilizer and response tensors are derived at each state without assigning an independent equation of motion to a group label. The next section uses that same response to determine what can be observed by incremental elastic waves.

\begin{figure}
	\centering
	\includegraphics[width=1.0\textwidth]{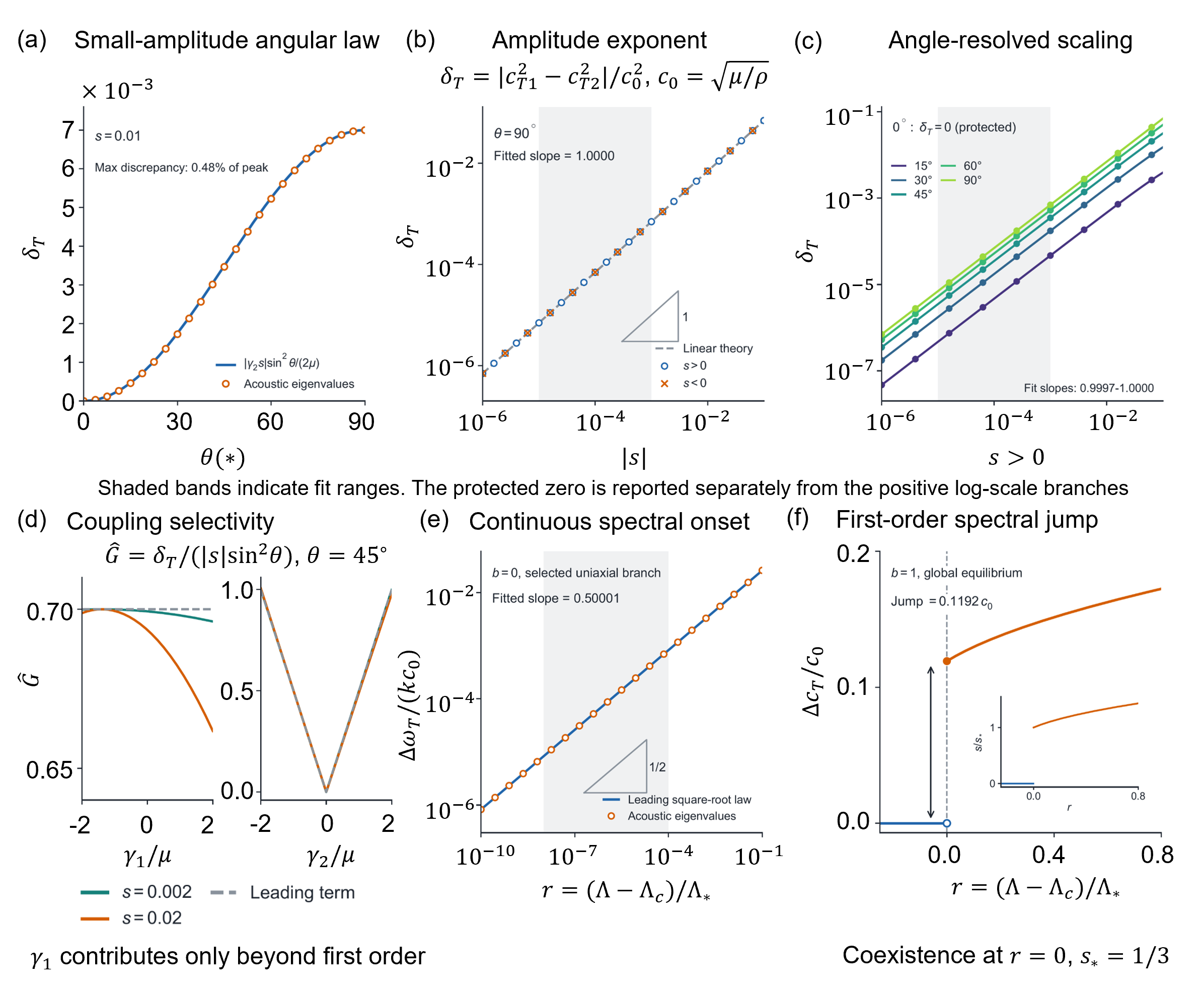}
	\caption{Scaling laws for symmetry-breaking spectral gaps.
		(a) Angular dependence of the transverse squared-speed gap $\delta_T$, showing agreement with the leading $\sin^2\theta$ law.
		(b) Linear scaling of $\delta_T$ with the structural amplitude $|s|$, with a fitted exponent of unity.
		(c) Angle-resolved amplitude scaling, demonstrating the protected axial degeneracy and linear gap opening at oblique directions.
		(d) Dependence of the normalized transverse gap coefficient on the constitutive couplings $\gamma_1$ and $\gamma_2$, confirming the leading-order selectivity to $\gamma_2$.
		(e) Continuous spectral onset along a selected uniaxial Landau branch, exhibiting the square-root scaling $\Delta\omega_T\propto(\Lambda-\Lambda_c)^{1/2}$.
		(f) Discontinuous transverse phase-speed gap at a first-order transition, corresponding to a finite jump in the equilibrium structural amplitude.
		Phase speeds are normalized by $c_0=\sqrt{\mu/\rho}$, and all quantities are dimensionless.}\label{fig5}
\end{figure}

\section{Spectral consequences and falsifiable predictions}
\label{sec:spectral_theory}

A wave experiment samples the stiffness on the time scale of its small-amplitude oscillations. Accordingly, the structural evolution of Section~\ref{sec:constitutive_theory} and the incremental probe must be distinguished. The central predictions below concern a frozen structural state during each oscillation and a base state that changes slowly between measurements. This gives a direct test of the constitutive dependence on \(\mathbf Q\). Relaxation during the oscillation introduces additional dynamic moduli, which are discussed separately so that a static symmetry signature is not confused with a kinetic dispersion effect.

\subsection{Incremental operator and acoustic covariance}
\label{sec:incremental_operator}

Small-amplitude elastic waves probe the tangent response of the current state. Let \(\mathcal Z_0\) be a locally homogeneous base state. Its complete symmetry is
\begin{equation}
 H_{\mathcal Z_0}=H_{\mathbf C_0}\cap H_{\mathbf Q_0}\cap H_{\boldsymbol\xi_0}\cap\cdots.
 \label{eq:base_state_symmetry}
\end{equation}
When the deformation and other variables preserve the structural symmetry, \(H_{\mathcal Z_0}=H_{\mathbf Q_0}\). The incremental equations are
\begin{equation}
 \rho_0\ddot u_i=\partial_j\left(\mathcal A_{0ijkl}\partial_lu_k\right),
 \label{eq:incremental_elastodynamics}
\end{equation}
where \(\mathbb A_0=\mathbb A_0(\mathcal Z_0)\) is the instantaneous spatial elasticity tensor.
At finite deformation this tensor is obtained from the derivative with respect to \(\mathbf F\), not by identifying the material tensor \(\mathbb C\) directly with a spatial modulus. Define \(W(\mathbf F)=\Psi(\mathbf F^T\mathbf F,\ldots)\) at fixed internal state. With \(J=\det\mathbf F\), the nominal tangent and its push-forward are
\begin{equation}
 \mathcal K_{iJkL}=\frac{\partial^2W}{\partial F_{iJ}\partial F_{kL}},
 \qquad
 \mathcal A_{0ijkl}=J^{-1}F_{jJ}F_{lL}\mathcal K_{iJkL}.
 \label{eq:spatial_tangent_pushforward}
\end{equation}
This construction includes the stress-dependent geometric terms contained in the \(\mathbf F\)-derivative. In Eq.~\eqref{eq:incremental_elastodynamics}, \(\rho_0\) denotes the mass density of the current homogeneous base state; if \(\rho_{\rm ref}\) is the referential density, then \(\rho_0=\rho_{\rm ref}/J\). At \(\mathbf F=\mathbf I\) and zero base stress, Eq.~\eqref{eq:spatial_tangent_pushforward} reduces to the small-strain modulus in Eq.~\eqref{eq:explicit_tangent_components} plus its isotropic part. At finite strain, referential symmetry operations must also be expressed in the current orthonormal frame. For a proper material rotation preserving \(\mathbf C_0\), the corresponding spatial rotation is \(\mathbf r=\mathbf F_0g\mathbf F_0^{-1}\). Below the symbol \(g\) in spatial covariance formulas denotes this spatial representative; in the undeformed example the reference and spatial representatives coincide.
For a plane wave \(\mathbf u=\mathbf p\exp[ik(\widehat{\mathbf k}\cdot\mathbf x-ct)]\), define
\begin{equation}
 \mathsf A_{ik}(\widehat{\mathbf k};\mathcal Z_0)
 =\mathcal A_{0ijkl}\widehat k_j\widehat k_l.
 \label{eq:acoustic_tensor}
\end{equation}
The polarization and phase speed solve
\begin{equation}
 \mathsf A(\widehat{\mathbf k};\mathcal Z_0)\mathbf p_n
 =\rho_0c_n^2\mathbf p_n.
 \label{eq:acoustic_eigenproblem}
\end{equation}

Substitution of the plane wave into Eq.~\eqref{eq:incremental_elastodynamics} cancels the common factor \(-k^2\), leaving this eigenproblem. For the conservative hyperelastic tangent, major symmetry makes \(\mathsf A\) real symmetric, so its polarizations can be chosen orthonormal. A stable propagation regime requires strong ellipticity, \(\mathbf p\cdot\mathsf A(\widehat{\mathbf k})\mathbf p>0\) for all nonzero \(\mathbf p\) and every unit propagation direction. This requirement guarantees positive squared speeds and is logically separate from parent covariance. Thus a symmetry-allowed constitutive law must still be tested for stability before its eigenvalues are interpreted as measurable sound speeds.
Parent invariance and constitutive differentiation imply, for \(g\in H_{\mathcal Z_0}\),
\begin{equation}
 \mathsf A(g^T\widehat{\mathbf k};\mathcal Z_0)
 =g^T\mathsf A(\widehat{\mathbf k};\mathcal Z_0)g.
 \label{eq:acoustic_covariance}
\end{equation}
Hence symmetry-related propagation directions have identical eigenvalue spectra.
Covariance follows by transforming both the propagation direction and the displacement polarization in the quadratic incremental energy. If \(\mathbf p_n\) solves the problem at \(\widehat{\mathbf k}\), then \(g^T\mathbf p_n\) solves it at \(g^T\widehat{\mathbf k}\) with the same eigenvalue. A constitutive model and a directional measurement can therefore be compared through complete rotated speed and polarization patterns, rather than a single gap. Since \(\mathsf A(-\widehat{\mathbf k})=\mathsf A(\widehat{\mathbf k})\) in this local elastic model, operations that reverse the propagation direction can also constrain its spectrum. The directed little group below is a sufficient symmetry group for the protection arguments; the group preserving the unoriented propagation line can impose additional restrictions.
The symmetry acting at a fixed propagation direction is the little group
\begin{equation}
 H_{\widehat{\mathbf k},\mathcal Z_0}
 =\{g\in H_{\mathcal Z_0}:g^T\widehat{\mathbf k}=\widehat{\mathbf k}\}.
 \label{eq:little_group}
\end{equation}
For every element of this group, the acoustic tensor commutes with the polarization representation. It is therefore block diagonal on the little-group representation sectors; a multidimensional irreducible sector can protect a modal degeneracy.

The precise protection statement concerns a symmetric operator on an irreducible real polarization sector. A nontrivial splitting would define invariant eigenspaces inside that sector, contradicting irreducibility; hence the operator must be scalar there. When equivalent sectors occur more than once, symmetry can allow coupling between their copies, so representation dimension alone does not determine every degeneracy. For the isotropic transverse plane used here there is one two-dimensional rotation sector, and the protection is unambiguous. Reducing the little group removes this restriction and allows a gap, while neither its magnitude nor its nonzero existence follows from subgroup reduction alone.

Figure~\ref{fig4} uses the same three states and coefficients as Fig.~\ref{fig3}, with $\rho_0=1$. The isotropic transverse surfaces coincide for every direction, while the uniaxial state retains the exact axial degeneracy. Biaxial states can possess isolated acoustic axes, but the positions of the illustrated crossings move under symmetry-preserving parameter changes. Polarization sectors are followed through crossings instead of labeling modes solely by speed; at degeneracy the combined eigenspace, rather than its individual basis vectors, is observable. Supplementary Section~\ref{sec:acoustics} gives the meridional block reduction and crossing diagnostics.

\begin{figure}
	\centering
	\includegraphics[width=1.0\textwidth]{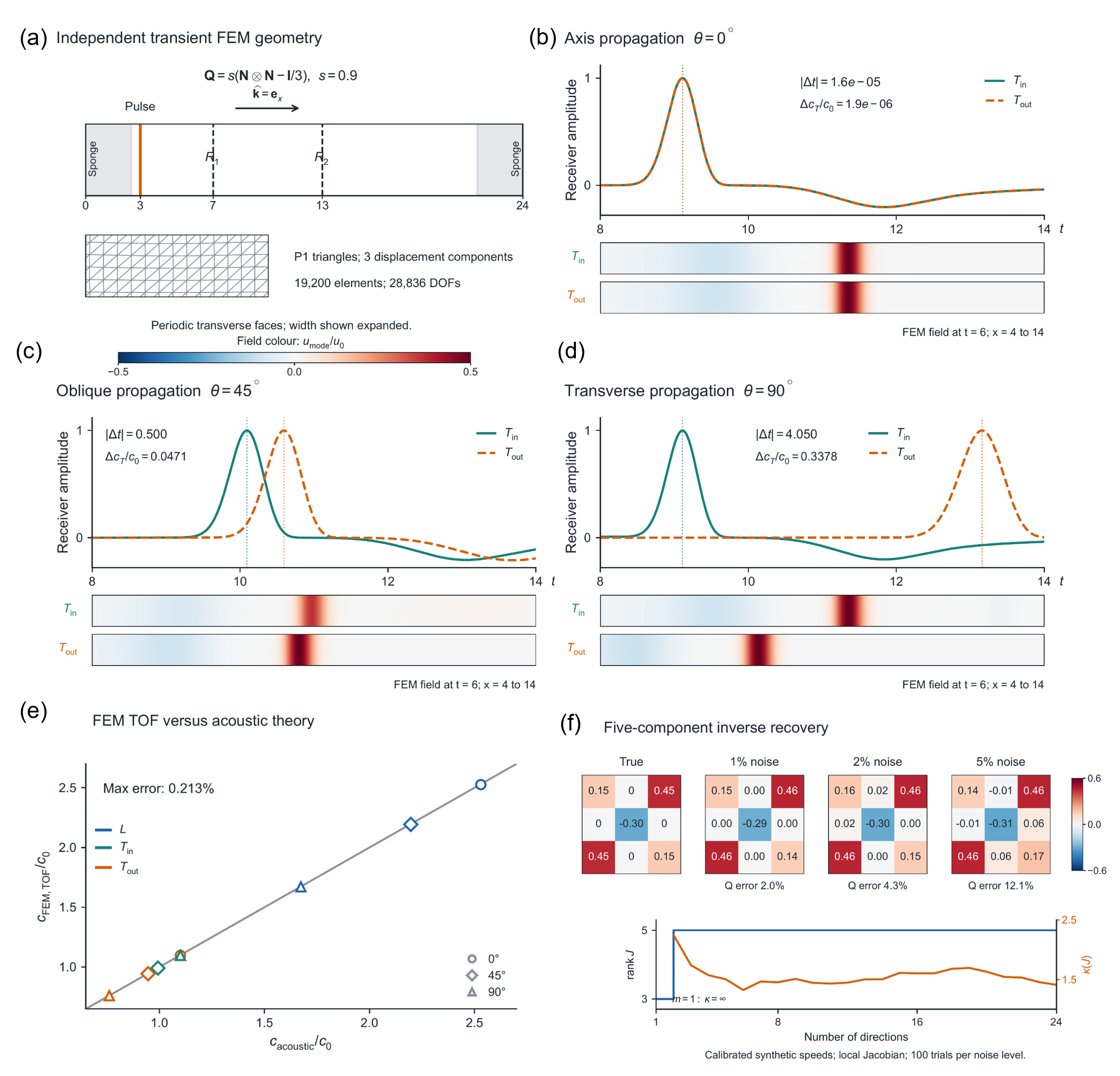}
	\caption{Full-field verification and identifiability of the structural symmetry state.
		(a) Independent transient finite-element model with periodic transverse boundaries and absorbing end regions.
		(b--d) Receiver waveforms and displacement fields for propagation angles of $0^\circ$, $45^\circ$, and $90^\circ$, demonstrating symmetry-protected transverse degeneracy and direction-dependent mode splitting.
		(e) Comparison of FEM time-of-flight pulse speeds with nondispersive acoustic-tensor phase-speed predictions, showing a maximum relative error of $0.213\%$.
		(f) Reconstruction of the five independent components of $\mathbf Q$ from synthetic acoustic measurements under different noise levels, together with the rank and conditioning of the inverse sensitivity matrix.
		Phase speeds are normalized by $c_0=\sqrt{\mu/\rho}$, and all quantities are dimensionless.}\label{fig6}
\end{figure}

\subsection{Degeneracy, symmetry breaking, and spectral gaps}
\label{sec:degeneracy_splitting}

At \(\mathbf Q=\mathbf0\), an undeformed isotropic solid has
\begin{equation}
 \mathsf A_{\rm iso}=(\lambda+\mu)\widehat{\mathbf k}\otimes\widehat{\mathbf k}+\mu\mathbf I.
 \label{eq:isotropic_acoustic_tensor}
\end{equation}
The longitudinal speed is \(c_L^2=(\lambda+2\mu)/\rho_0\), while the transverse plane is degenerate:
\begin{equation}
 c_{T1}=c_{T2}=\sqrt{\mu/\rho_0}.
 \label{eq:isotropic_transverse_degeneracy}
\end{equation}
This equality is protected by the axial \(SO(2)\) little group acting irreducibly on the transverse plane.
Near the isotropic state, smooth constitutive dependence gives
\begin{equation}
 \mathbb A_0(\mathbf Q)=\mathbb A_{\rm iso}+\mathbb B[\mathbf Q]+O(\|\mathbf Q\|^2),
 \qquad
 \mathsf A=\mathsf A_0+\delta\mathsf A+O(\|\mathbf Q\|^2).
 \label{eq:tangent_perturbation}
\end{equation}
Let \(\mathbf e_1,\mathbf e_2\) be an orthonormal basis of the isotropic transverse subspace. For the uniaxial state \(\mathbf Q=s(\mathbf N\otimes\mathbf N-\mathbf I/3)\), define
\begin{equation}
 K_{ab}^{(T)}=\mathbf e_a\cdot\mathsf A_1(\widehat{\mathbf k},\mathbf N)\mathbf e_b,
 \qquad a,b=1,2,
 \label{eq:transverse_projected_matrix}
\end{equation}
where \(\delta\mathsf A=s\mathsf A_1+O(s^2)\). If \(\eta_\pm\) are the two eigenvalues of \(K^{(T)}\), degenerate perturbation theory yields
\begin{equation}
 \rho_0c_{T\pm}^2=\mu+s\eta_\pm+O(s^2),
 \label{eq:shear_speed_split}
\end{equation}
and, for fixed wavenumber,
\begin{equation}
 \Delta\omega_T=|\omega_{T+}-\omega_{T-}|
 =\frac{k|s|}{2\rho_0c_T}|\eta_+-\eta_-|+O(k s^2).
 \label{eq:frequency_split}
\end{equation}
Thus symmetry reduction permits splitting but does not require it for every direction. When \(\widehat{\mathbf k}\parallel\mathbf N\), axial rotations remain in the little group and protect the transverse degeneracy. For a generic oblique direction the little group is smaller and splitting is generally allowed. A biaxial \(D_2\) state has no continuous axial protection for a generic direction.
More generally, for branches \(m,n\) that are degenerate at the high-symmetry state,
\begin{equation}
 \Delta_{mn}=|\omega_m-\omega_n|
 =\alpha_{mn}(\widehat{\mathbf k})\|\mathbf Q\|^p+o(\|\mathbf Q\|^p),
 \label{eq:gap_scaling}
\end{equation}
For this expansion, take a fixed normalized structural path $\mathbf Q=\varepsilon\widehat{\mathbf Q}$ and an analytic acoustic perturbation with a nonzero finite Taylor coefficient. The exponent \(p\) is the lowest nonzero symmetry-allowed order on that path; the prefactor can depend on $\widehat{\mathbf Q}$ as well as propagation direction. A smooth non-analytic law need not have a finite power-law onset. A linear coupling gives \(p=1\); residual symmetry or a vanishing constitutive coefficient can suppress that term and produce \(p\geq2\). The angular zeros of \(\alpha_{mn}\) are as informative as its nonzero values because symmetry-protected and accidental degeneracies have different robustness.

An explicit angular law can now be derived from the same minimal potential, rather than introduced as an independent spectral ansatz. Put \(\mathbf m=\widehat{\mathbf k}\), \(q_m=\mathbf m\cdot\mathbf Q\mathbf m\), and consider the undeformed, stress-free state with frozen \(\mathbf Q\). Contracting Eq.~\eqref{eq:explicit_tangent_components} with \(m_jm_l\) yields
\begin{equation}
 \delta\mathsf A=\gamma_1\big[\mathbf m\otimes(\mathbf Q\mathbf m)
 +(\mathbf Q\mathbf m)\otimes\mathbf m\big]
 +\frac{\gamma_2}{2}\big[\mathbf Q+q_m\mathbf I
 +(\mathbf Q\mathbf m)\otimes\mathbf m+\mathbf m\otimes(\mathbf Q\mathbf m)\big].
 \label{eq:explicit_acoustic_perturbation}
\end{equation}
On the transverse plane, \(P_T=\mathbf I-\mathbf m\otimes\mathbf m\) eliminates every term containing an exterior factor \(\mathbf m\), leaving
\begin{equation}
 P_T\delta\mathsf A P_T\big|_T
 =\frac{\gamma_2}{2}\big(P_T\mathbf QP_T+q_mP_T\big)\big|_T.
 \label{eq:explicit_transverse_block}
\end{equation}
The \(\gamma_1\) coupling therefore makes no first-order contribution to transverse splitting, although it can mix longitudinal and transverse polarizations and affect higher-order corrections. This provides a concrete example of a structurally anisotropic state whose particular measured gap can be insensitive to one admissible coupling.
For \(\mathbf Q=s(\mathbf N\otimes\mathbf N-\mathbf I/3)\), write \(\cos\vartheta=\mathbf m\cdot\mathbf N\). Away from the axis, choose \(\mathbf e_1=P_T\mathbf N/\sin\vartheta\) and \(\mathbf e_2=(\mathbf m\times\mathbf N)/\sin\vartheta\). These directions diagonalize the first-order transverse block, with
\begin{equation}
 \eta_1=\frac{\gamma_2}{6},\qquad
 \eta_2=\frac{\gamma_2}{2}\left(\cos^2\vartheta-\frac23\right),
 \qquad \eta_1-\eta_2=\frac{\gamma_2}{2}\sin^2\vartheta.
 \label{eq:explicit_transverse_eigenvalues}
\end{equation}
Here the subscripts identify polarization branches rather than sorting them by speed, since the ordering depends on the sign of \(s\gamma_2\). The corresponding observable gap is
\begin{equation}
 \Delta\omega_T=\frac{k|\gamma_2|\,|s|}{4\rho_0c_T}\sin^2\vartheta
 +O(k s^2),\qquad
 |c_{T1}^2-c_{T2}^2|=\frac{|\gamma_2|\,|s|}{2\rho_0}\sin^2\vartheta+O(s^2).
 \label{eq:explicit_angular_gap}
\end{equation}
The expansion assumes a nonzero longitudinal--transverse separation in the isotropic reference state. The transverse basis is undefined at \(\vartheta=0\), but the gap tends to zero there and axial symmetry protects the exact degeneracy. At oblique directions, the two transverse polarizations acquire longitudinal admixtures at higher order and are properly described as quasi-transverse modes. The angular law is consequently a controlled small-order prediction of the chosen coupling, while the axial zero is a symmetry statement that survives additional symmetry-compatible constitutive terms.
Combining Eq.~\eqref{eq:explicit_angular_gap} with the selected continuous branch in Eq.~\eqref{eq:continuous_order_scaling}, and evaluating smooth coefficients at the transition, gives
\begin{equation}
 \Delta\omega_T=\frac{k|\gamma_2|}{4\rho_0c_T}
 \sqrt{\frac{3a_0}{2c}}\,\sin^2\vartheta\,(\Lambda-\Lambda_c)^{1/2}
 +o\big(k(\Lambda-\Lambda_c)^{1/2}\big).
 \label{eq:control_parameter_gap}
\end{equation}
This square-root onset needs a continuous, unbiased transition, a selected uniaxial branch, and a nonzero \(\gamma_2\). It is not a universal exponent for all symmetry transitions. A first-order branch instead produces a finite change of the spectral response when the selected equilibrium state jumps, provided the relevant coupling and geometry are sensitive to that change. A general gap with leading power \(p\) scales as \((\Lambda-\Lambda_c)^{p/2}\) along this mean-field continuous path. These qualifications allow an observed deviation to distinguish a different constitutive coupling or transition mechanism from a failure of the covariance principle itself.

Figure~\ref{fig5} compares the leading laws with the full acoustic eigenvalues. The squared-speed gap is $\delta_T=|c_{T1}^2-c_{T2}^2|/c_0^2$, while the speed gap is $g_T=|c_{T1}-c_{T2}|/c_0$; these are different observables. The reported small-amplitude angular discrepancy is approximately $0.48\%$ of the first-order peak at $s=0.01$. At the equator, the shear eigenvalues of the present model are exactly $(\mu+\gamma_2s/6)/\rho_0$ and $(\mu-\gamma_2s/3)/\rho_0$, giving an exactly linear squared-speed gap within the stable shear-branch regime. For $\gamma_2=0$, longitudinal--transverse mixing instead produces an oblique $O(s^2)$ gap. The continuous and first-order transition plots therefore test a selected constitutive branch, with the exact formulas and normalizations collected in Supplementary Section~\ref{sec:gaps}.

The forward chain established by the theory is
\begin{equation}
	\mathcal{F}_{\mathrm{wave}}:
	\mathbf Q
	\mapsto
	\bigl(H_{\mathbf Q},\,\mathbb A_0(\mathbf Q),\,
	\mathsf A(\widehat{\mathbf k};\mathbf Q),\,
	\{c_n,\mathbf p_n\}\bigr).
	\label{eq:forward_chain}
\end{equation}
The inverse direction is conditional. A measured splitting establishes only that the complete experimental state and propagation geometry do not protect the former degeneracy. Unique recovery of \(\mathbf Q\) additionally requires a calibrated constitutive map, enough propagation directions, and control of prestress, damage, and other symmetry-bearing variables. Accordingly, the framework predicts symmetry-compatible spectral patterns, not universal numerical gaps.

Within these limits, four observations are directly testable: The isotropic state has two degenerate transverse modes for every propagation direction. A uniaxial state retains transverse degeneracy along its symmetry axis but generally permits splitting away from that axis. Near a continuous transition, an allowed gap follows Eq.~\eqref{eq:gap_scaling}; a first-order jump in \(\mathbf Q\) produces a corresponding discontinuity in an otherwise smooth spectral observable. States related by a parent rotation have rotated slowness and polarization patterns with identical eigenvalues at symmetry-related directions, as required by Eq.~\eqref{eq:acoustic_covariance}.

These statements separate the present result from a generic assertion that ``anisotropy splits waves.'' The predicted content is the covariance of the full directional spectrum, the directions on which degeneracy remains protected, and the scaling order by which an allowed gap emerges.

A conditional reconstruction is possible within the explicit small-order model. If the director is known independently and \(\gamma_2\) has been calibrated, polarization-labeled transverse speeds give
\begin{equation}
 s=\frac{2\rho_0}{\gamma_2\sin^2\vartheta}
 (c_{T1}^2-c_{T2}^2)+O(s^2),
 \qquad \gamma_2\sin^2\vartheta\neq0.
 \label{eq:conditional_amplitude_recovery}
\end{equation}
The sign is available only when the polarization branches are identified; an unsigned gap gives \(|s\gamma_2|\). Measurements close to the axis are poorly conditioned because the sensitivity vanishes as \(\sin^2\vartheta\), even though they are useful for testing the protected zero. If neither the structural amplitude nor the coupling coefficient is known, the leading splitting identifies their product and cannot determine them separately. Multiple propagation directions and vector-resolved polarizations help locate the director and test the angular law, while an independent structural measurement is needed to establish that a change in speed originates from the proposed structural state.

For a biaxial state or a general calibrated potential, this identifiability requirement can be stated locally. Expand \(\mathbf Q=\sum_{A=1}^5q_A\mathbf T_A\) in an orthonormal basis of \(\operatorname{Sym}_0(3)\), and let \(y_\ell\) be the collection of measured spectral quantities at selected directions. Their sensitivities obey
\begin{equation}
 \delta y_\ell=\sum_{A=1}^5\mathcal J_{\ell A}\delta q_A,
 \qquad \mathcal J_{\ell A}=\frac{\partial y_\ell}{\partial q_A},
 \qquad \operatorname{rank}\mathcal J=5.
 \label{eq:inverse_local_rank}
\end{equation}
Full column rank is a local identifiability condition at a regular state with all nuisance parameters fixed; it does not establish global uniqueness or robust recovery under noise. At exact degeneracy, individual polarization eigenvectors are not unique, and derivatives should be formulated using the degenerate eigenspace or its projector. This makes clear why a single split pair is insufficient to identify a general five-component structural tensor.

The complete tangent identity also gives a speed-only sufficient reconstruction criterion for the explicit model. Taking the trace of the acoustic tensor yields the exact relation
\begin{equation}
 z(\mathbf m):=\rho_0\sum_{n=1}^3c_n^2(\mathbf m)-(\lambda+4\mu)
 =a_V\,\mathbf m\cdot\mathbf Q\mathbf m.
 \label{eq:acoustic_trace_recovery}
\end{equation}
Unlike individual polarization labels, the sum is unchanged by branch permutation or a transverse degeneracy. For $\mathbf Q=\sum_Aq_A\mathbf T_A$, directional data satisfy
\begin{equation}
 z_m=a_V\sum_{A=1}^5B_{mA}q_A,\qquad
 B_{mA}=\mathbf m_m\cdot\mathbf T_A\mathbf m_m.
 \label{eq:trace_design}
\end{equation}
If $a_V\neq0$ and $\operatorname{rank}\mathbf B=5$, this linear system recovers $\mathbf Q$ uniquely within the calibrated model. Five suitable directions suffice; for example $\mathbf e_1$, $\mathbf e_2$, $(\mathbf e_1+\mathbf e_2)/\sqrt2$, $(\mathbf e_1+\mathbf e_3)/\sqrt2$, and $(\mathbf e_2+\mathbf e_3)/\sqrt2$ determine the two independent diagonal and three off-diagonal components. When $a_V=0$, this trace observable is blind even though the exact symmetry result established above still applies. This is a sufficient observation design, not a minimum-direction theorem for all modal data.

When density, prestress, or coupling coefficients are estimated jointly, the five-column rank test alone is insufficient. Let $\boldsymbol\nu$ collect these nuisance parameters and let the noise covariance be $\boldsymbol\Sigma$. Define whitened sensitivities $\widetilde{\mathcal J}_Q=\boldsymbol\Sigma^{-1/2}\mathcal J_Q$, $\widetilde{\mathcal J}_\nu=\boldsymbol\Sigma^{-1/2}\mathcal J_\nu$, and $P_\nu=\widetilde{\mathcal J}_\nu\widetilde{\mathcal J}_\nu^+$, with $+$ denoting the Moore--Penrose inverse. The structural information remaining after nuisance variations is
\begin{equation}
 \mathcal I_{Q|\nu}=\widetilde{\mathcal J}_Q^T
 (\mathbf I-P_\nu)\widetilde{\mathcal J}_Q.
 \label{eq:nuisance_information}
\end{equation}
Positive definiteness is the corresponding regular, first-order local distinguishability condition. A null direction can be mimicked by nuisance changes; small eigenvalues imply noise amplification. In particular, the elastic response is invariant under $\mathbf Q\mapsto r\mathbf Q$, $\gamma_i\mapsto\gamma_i/r$ for nonzero $r$, so acoustic data alone cannot fix the structural normalization and both unknown couplings simultaneously. Independent calibration or structural measurements must remove this ambiguity. This limitation persists even with dense angular sampling.

\subsection{Full-field verification and conditional identification}
\label{sec:verification}

Figure~\ref{fig6}(a--e) compares the acoustic predictions with a separately integrated transient displacement field. The benchmark uses a homogeneous periodic two-dimensional strip with three displacement components, a stiffness Hessian obtained from the same scalar energy, and absorbing end regions. It tests bulk propagation, rather than free-surface plate modes. For the three tested propagation angles and nine angle--branch combinations, the reported maximum pulse-speed discrepancy is $0.2133\%$. Mesh refinement reduces the equatorial discrepancy from $1.3389\%$ to $0.2133\%$. Agreement verifies the numerical implementation of this constitutive example; it does not establish a material-specific constitutive calibration. In anisotropic elasticity, the group-velocity vector need not be parallel to the wavevector even without frequency dispersion. The present comparison concerns the axial transit of a transversely uniform periodic-strip pulse; finite-aperture beam measurements require a separate ray and group-velocity analysis.

The separate synthetic inverse test in Fig.~\ref{fig6}(f) fits all five tensor components from three speeds at each of 24 directions, with density and elastic couplings fixed. The reported relative-sensitivity matrix has rank five and condition number $1.411$. Median tensor errors are $2.041\%$, $4.275\%$, and $12.052\%$ for relative speed-noise levels of $1\%$, $2\%$, and $5\%$, respectively, over 100 realizations per level. The data are generated by the calibrated acoustic model, not by the transient FEM benchmark. This is a conditional recovery test under a matched forward model. Noisy biaxiality in the estimate is not evidence of a physical symmetry change. Supplementary Section~\ref{sec:fem_inverse} gives the FEM equations, fitting objective, and noise protocol.

The spectral result assumes a locally homogeneous base state, wavelengths long compared with the microstructural scale, and small-amplitude increments about that state. In classical local elasticity, \(\omega_n=c_n(\widehat{\mathbf k})k\); the relevant object is therefore a directional acoustic or slowness surface, not a frequency-dispersive band structure. Prestress and other internal variables must be included in \(H_{\mathcal Z_0}\). Finally, the second-order tensor \(\mathbf Q\) resolves the hierarchy \(SO(3)\rightarrow O(2)\rightarrow D_2\), but it is not a complete descriptor of cubic, chiral, or more complicated material symmetries.

More quantitatively, if \(\ell_{\rm micro}\) is a representative microstructural length and \(L_{\rm state}\) the length over which the base state varies appreciably, a local plane-wave interpretation requires \(\ell_{\rm micro}\ll2\pi/k\ll L_{\rm state}\). The state must also be effectively frozen, \(\omega\tau_Q\gg1\), in the structural sectors coupled to the probe, while the base state should remain approximately constant over the acquisition interval. Temperature is treated as fixed for the explicit model; when thermoelastic coupling is significant, the relevant tangent must instead be chosen consistently with the thermal time scale, for example an adiabatic tangent. These scale choices specify which constitutive response an experiment actually samples.

Eliminating linearly relaxing structural increments generally produces a complex, frequency-dependent tangent. For the particular quadratic-in-strain coupling in Eq.~\eqref{eq:minimal_energy}, however, $\Psi_{,\mathbf Q\mathbf E}=\mathbf0$ at zero base strain. The wave probe then does not drive a first-order structural increment, and the frozen conservative acoustic model is also the exact linear mechanical response about a stationary stress-free state. The condition $\omega\tau_Q\gg1$ is sufficient in the more general coupled setting, but unnecessary for this decoupled linearization. The dynamic Schur-complement formula, its stability assumptions, and the effect of structural zero modes are given in Supplementary Section~\ref{sec:dynamic}.

\section{Conclusions}
\label{sec:conclusions}

An evolving structural tensor can select residual material symmetry within one smooth parent-invariant constitutive law. For the minimal space $\operatorname{Sym}_0(3)$, eigenvalue multiplicities distinguish isotropic, uniaxial, and biaxial states, while the stabilizer distinguishes a change of symmetry type from reorientation within one orbit. Continuity belongs to the state and its response tensors; subgroup labels can change at repeated-eigenvalue strata. The construction therefore does not require an independent dynamics for a symmetry-group label.
The mechanical content depends on how the state enters the energy. Parent invariance guarantees symmetry inheritance, whereas equality of structural and tangent symmetry requires a coupling that separates states. For the explicit two-coupling potential, the stiffness-contraction reconstruction proves this equality for every nonzero coupling pair. The same potential generates admissible structural kinetics, a positive-energy sufficient bound, and the elastic tangent, making their mutual consistency explicit. Exact isotropy nevertheless remains invariant under unbiased deterministic kinetics; seeded ordering and externally driven class changes must be distinguished from an unforced evolution.
The acoustic consequences separate symmetry constraints from constitutive magnitudes. The isotropic transverse pair is protected in every direction, and a uniaxial state retains its exact axial degeneracy. Away from that axis, the selected potential gives a leading squared-speed gap proportional to $|\gamma_2s|\sin^2\vartheta$. Suppressing this coupling can shift the leading opening to second order. A square-root onset with a control parameter follows only along a selected continuous mean-field branch; it is not a universal consequence of symmetry reduction. The finite-amplitude spectra and supplied transient FEM benchmark are consistent with these forward predictions, with a reported maximum pulse-speed discrepancy of $0.2133\%$.
Observability is more restrictive than forward covariance. The directional acoustic trace provides an exact linear reconstruction under a stated calibrated coupling and a rank-five direction design. General modal recovery additionally depends on noise, degeneracies, and nuisance sensitivities; a measured gap alone cannot establish the structural origin of anisotropy. The synthetic inversion demonstrates conditional recovery under a matched model, not global uniqueness or experimental validation.
The present results establish a tractable constitutive-to-acoustic route for second-order structural symmetry. Material-specific use requires a measurable interpretation of $\mathbf Q$, independent calibration of the couplings and kinetics, and wave measurements consistent with the deformation, thermal, and structural time scales. Higher-order descriptors or tensor sets are needed for symmetries outside the present hierarchy. These extensions should retain the distinction between state symmetry, mechanically expressed symmetry, and the information accessible to a particular experiment.

\section*{Declaration of competing interest}
The author declares that he has no known competing financial interests or personal relationships that could have appeared to influence the work reported in this paper.

\section*{Acknowledgments}
This work was supported by the Adolf Martens Fellowship (Grant No. BAM-AMF-2025-1).

\section*{Data availability}
Data will be made available on request

\appendix
\section{Invariant geometry and the prescribed state path}
\label{sec:geometry}
Figures~\ref{fig1}(a--c) give a concrete representation of this invariant classification. The ellipsoids in panel (a) have positive principal radii \(\exp(1.4q_i)\), obtained from \(\exp(1.4\mathbf Q)\); they visualize the structural spectrum rather than a deformation or a measured specimen shape. Since the exponential is injective on the real eigenvalues, it preserves their multiplicities and the stabilizer, while \(\prod_i\exp(1.4q_i)=1\). The invariant domain in panel (b) follows from the discriminant of the characteristic polynomial in Supplementary Section~\ref{app:orbit_geometry},
\begin{equation}
	\operatorname{disc}(\chi_{\mathbf Q})=\prod_{i<j}(q_i-q_j)^2
	=\frac12J_2^3-3J_3^2
	=\frac12J_2^3\beta_Q^2\geq0.
	\label{eq:figure_discriminant}
\end{equation}
Thus the boundary is precisely the repeated-eigenvalue locus, whereas panel (c) resolves biaxiality within the interior. At \(J_3=0\) and \(J_2>0\), the normalized biaxiality reaches unity; at the isotropic origin it remains undefined. This geometry distinguishes an amplitude approaching zero from a nonzero state approaching uniaxiality.

The eigenvalue path in Fig.~\ref{fig1}(f) illustrates a change of stratum without introducing a discontinuity in the tensor itself. In the plotting units, it is prescribed by
\begin{equation}
	\frac{(q_1,q_2,q_3)}{s_*}
	=\left(\frac{2u}{3},-\frac{u}{3}+d,-\frac{u}{3}-d\right),
	\qquad
	u=\sqrt{\frac{\max(r,0)}{2}},\quad
	d=0.22\sqrt{\max(r-0.65,0)}.
	\label{eq:figure_state_path}
\end{equation}
Here \(r\) is the dimensionless path parameter and \(s_*=1/3\). The origin, the segment with \(u>0,d=0\), and the segment with three distinct eigenvalues have stabilizers \(SO(3)\), \(O(2)\), and \(D_2\), respectively. This is a prescribed geometric path, not a claim that the unloaded quartic potential alone stabilizes its biaxial segment. A dynamically generated class change is examined separately in Fig.~\ref{fig2}(c).

\section{Stiffness evaluation, covariance, and stability diagnostics}
\label{sec:stiffness}
Figure~\ref{fig3} evaluates these constitutive consequences using \(\lambda=2\), \(\mu=1\), \(\gamma_1=0.6\), and \(\gamma_2=1.4\). The three structural states in panels (a--c) are \(\mathbf Q=\mathbf0\), \(\operatorname{diag}(-0.3,-0.3,0.6)\), and \(\operatorname{diag}(-0.55,0.1,0.45)\), respectively. All matrices are expressed in the structural eigenframe with engineering strain components \((E_{11},E_{22},E_{33},2E_{23},2E_{13},2E_{12})\). For the uniaxial state, the equalities \(C^V_{11}=C^V_{22}\), \(C^V_{13}=C^V_{23}\), \(C^V_{44}=C^V_{55}\), and \(C^V_{66}=(C^V_{11}-C^V_{12})/2\) exhibit transverse isotropy directly. The biaxial state permits an orthotropic matrix structure, although the chosen low-parameter potential does not span all nine independent orthotropic moduli. An independent numerical Hessian of the scalar mechanical energy agrees with the analytical stiffness to within \(8.88\times10^{-16}\).

The directional surfaces in Fig.~\ref{fig3}(d) use the stiffness associated with a prescribed longitudinal strain,
\begin{equation}
	M(\mathbf m)=(\mathbf m\otimes\mathbf m):\mathbb C:(\mathbf m\otimes\mathbf m)
	=\lambda+2\mu+2(\gamma_1+\gamma_2)\,\mathbf m\cdot\mathbf Q\mathbf m,
	\qquad |\mathbf m|=1.
	\label{eq:figure_directional_stiffness}
\end{equation}
Its normalized radius \(M/(\lambda+2\mu)\) changes from a sphere to an axisymmetric surface and then to a surface with three distinct principal directions. The shared normalization makes amplitude differences comparable across states. This quantity is \(\mathbf m\cdot\mathsf A(\mathbf m)\mathbf m\), rather than generally \(\rho_0c_L^2\): away from a principal direction, the quasi-longitudinal polarization need not coincide with \(\mathbf m\).

Figure~\ref{fig3}(e) tests parent covariance by rotating the state as well as the tensor components. For \(\mathbf Q'=\mathbf R^T\mathbf Q\mathbf R\), the diagnostic is
\begin{equation}
	\varepsilon_{\rm cov}
	=\frac{\|\mathbb C(\mathbf Q')-\mathbf R\star\mathbb C(\mathbf Q)\|_F}
	{\|\mathbb C(\mathbf Q)\|_F},\qquad
	(\mathbf R\star\mathbb C)_{ijkl}
	=R_{pi}R_{qj}R_{rk}R_{sl}C_{pqrs}.
	\label{eq:figure_stiffness_covariance}
\end{equation}
Using the complete fourth-order tensor avoids the coordinate-dependent Euclidean norm of an engineering Voigt matrix. Over 600 random proper rotations per state, the largest residual is approximately \(1.59\times10^{-15}\), consistent with covariance at floating-point precision.

The stability maps in Fig.~\ref{fig3}(f) distinguish two conditions that covariance alone cannot establish. Their color represents the acoustic margin
\begin{equation}
	\begin{aligned}
		m_{\rm ac}&=\min_{|\mathbf m|=1}\lambda_{\min}[\mathsf A(\mathbf m)],
		&m_{\rm en}&=\lambda_{\min}(\mathbb C_K),\\
		\mathbb C_K&=\mathbf D_K\mathbb C_V\mathbf D_K,
		&\mathbf D_K&=\operatorname{diag}(1,1,1,\sqrt2,\sqrt2,\sqrt2).
	\end{aligned}
	\label{eq:figure_stability_margins}
\end{equation}
A discrete angular minimum can overestimate the true minimum over the sphere; positive sampled values alone do not certify strong ellipticity. The Kelvin basis is orthonormal, so \(m_{\rm en}>0\) is equivalent to positive strain energy, while \(m_{\rm ac}>0\) establishes strong ellipticity. The dashed sufficient bound from Eq.~\eqref{eq:sufficient_elastic_stability} lies inside the stable region and becomes more restrictive as \(|s|\) increases. States outside that bound can remain stable: the three illustrated Kelvin margins are approximately \(2\), \(1.16\), and \(0.7793\), and the uniaxial acoustic margin is \(0.58\). These results also show why the sufficient bound must not be interpreted as the exact stability boundary.

\section{Structural-field evolution and orientation coarsening}
\label{sec:kinetics}
Figure~\ref{fig2}(a) integrates Eq.~\eqref{eq:explicit_Q_evolution} for the full three-dimensional tensor on a two-dimensional periodic domain. The dimensionless parameters are \(a=-1\), \(b=3\), \(c=2\), and \(\kappa_Q=\Gamma_Q=1\), with zero mechanical bias. A small random symmetric-traceless initial field is evolved on a \(256\times256\) grid using a Fourier pseudospectral Laplacian and second-order exponential time stepping. With \(t_Q=(\Gamma_Q|a|)^{-1}\), the snapshots at \(t/t_Q=0,1,10,100\) resolve amplitude growth, transient biaxial regions, and subsequent orientational organization. The ordered bulk approaches the uniaxial amplitude \(s_{\rm eq}=(3+\sqrt{57})/8\), but defects and spatial variation prevent the whole field from being assigned one exact \(O(2)\) stabilizer. Panel (d) checks the same unforced evolution against
\begin{equation}
	\mathcal D_Q^{\rm tot}=\Gamma_Q\int_{\mathcal B_0}|\mathbf A_Q^0|^2\,\mathrm dV,
	\qquad \frac{\mathrm d\mathcal F}{\mathrm dt}=-\mathcal D_Q^{\rm tot},
	\qquad \varepsilon_{\rm tr}=\max_{\mathbf X}|\operatorname{tr}\mathbf Q(\mathbf X,t)|.
	\label{eq:figure_dissipation_diagnostic}
\end{equation}
The numerical energy decreases and the trace residual remains of order \(10^{-16}\). Tensor contractions include the factor of two carried by off-diagonal symmetric components, so the dissipation uses the Frobenius norm of the actual tensor field.

The distinction between orbit motion and class change is tested by different protocols in Fig.~\ref{fig2}(b,c). Panel (b) constrains \(\mathbf Q=s_{\rm eq}(\mathbf n\otimes\mathbf n-\mathbf I/3)\) and uses positive-semidefinite mobility \(\Gamma_Q\mathbb P_{\rm orb}\), where \(\mathbb P_{\rm orb}\) projects onto the orbit tangent space. The eigenvalues and local symmetry type are consequently fixed while the orientation texture coarsens. Its characteristic length is extracted from the connected structure factor,
\begin{equation}
	\mathbf P(\mathbf X)=\frac{\mathbf Q(\mathbf X)}{\sqrt{J_2}}
	-\left\langle\frac{\mathbf Q}{\sqrt{J_2}}\right\rangle,
	\qquad S_P(\mathbf q)=\|\widehat{\mathbf P}(\mathbf q)\|_F^2,
	\qquad L_n=2\pi\frac{\sum_{\mathbf q\neq0}S_P(\mathbf q)}
	{\sum_{\mathbf q\neq0}|\mathbf q|S_P(\mathbf q)}.
	\label{eq:figure_texture_length}
\end{equation}
The observed increase in \(L_n\) demonstrates coarsening on a uniaxial orbit, without assigning a universal growth exponent or identifying this constrained flow with the unconstrained scalar-mobility quench. In panel (c), a homogeneous initially isotropic state is first seeded while \(a\) is reduced from \(0.6\) to \(-1\), and is subsequently subjected to a transverse symmetry-breaking strain bias through Eq.~\eqref{eq:mechanical_structural_force}. The eigenvalues then exhibit the sequence \(SO(3)\), \(O(2)\), and \(D_2\). The seed is required because the exact zero state cannot leave itself under the unbiased deterministic equation. The time-dependent controls supply power, so panel (c) is not a test of unforced energy decay.

\section{Coexistence interfaces and structural relaxation spectra}
\label{sec:interfaces}
Figures~\ref{fig2}(e,f) test the amplitude length scale and the structural relaxation spectrum independently. At isotropic--uniaxial coexistence, a fixed director reduces the interface problem to
\begin{equation}
	\begin{aligned}
		\kappa_Qs''&=a_ts-\frac b3s^2+\frac{2c}{3}s^3,
		&s(x)&=\frac{s_t}{1+\exp[-(x-x_0)/\ell_Q]},\\
		\ell_Q&=\sqrt{\frac{\kappa_Q}{a_t}},
		&a_t&=\frac{b^2}{27c},\qquad s_t=\frac b{3c}.
	\end{aligned}
	\label{eq:figure_coexistence_interface}
\end{equation}
The boundary-value solutions in panel (e) are computed independently of this analytical profile. Defining \(\ell_Q=(x_{90}-x_{10})/(2\ln9)\), where \(s(x_{10})=0.1s_t\) and \(s(x_{90})=0.9s_t\), gives the predicted square-root dependence on \(\kappa_Q\). The restoring curvature is the positive amplitude curvature \(\mu_Q=a_t\); it is not the zero curvature of a pure orientational mode. In panel (f), growth rates extracted from the Fourier amplitudes of small perturbations agree with Eq.~\eqref{eq:structural_linear_relaxation}. The tested curvatures include \(h_Q=-1\) for an unstable isotropic state, \(h_Q=0.5\) for a stable isotropic state, and \(h_Q=0\) and \(s_{\rm eq}(4cs_{\rm eq}-b)/3\) for orientation and amplitude perturbations of the ordered state. These are growth and relaxation rates of the structural variable, rather than elastic-wave frequencies.

\section{Normalized Landau branches used in the figures}
\label{sec:landau}
Figures~\ref{fig1}(d,e) compare the free-energy landscapes and their selected global minima, making the distinction between continuous onset and first-order coexistence explicit. With \(c=b_{\rm ref}=1\), introduce \(s_*=b_{\rm ref}/(3c)\), \(a_*=b_{\rm ref}^2/(27c)\), \(\Psi_*=b_{\rm ref}^4/(729c^3)\), and \(u=s/s_*\). The plotted sections are
\begin{equation}
	\frac{\Psi_Q}{\Psi_*}=
	\begin{cases}
		u^4-ru^2, & b=0,\quad a=-a_*r,\\
		u^4-2u^3+(1-r)u^2, & b=b_{\rm ref},\quad a=a_*(1-r).
	\end{cases}
	\label{eq:figure_scaled_landau}
\end{equation}
Here \(r\) is measured from the equilibrium transition of each case. For the selected nonnegative uniaxial branch, \(u_{\rm eq}=\sqrt{r/2}\) in the continuous case and \(u_{\rm eq}=[3+\sqrt{1+8r}]/4\) on the ordered side of the first-order case. At \(r=0\), the latter potential has equal minima at \(u=0\) and \(u=1\), separated by a barrier of height \(\Psi_*/16\). Panel (e) therefore displays an equilibrium amplitude jump, rather than a continuous sequence of stable intermediate states. The plotted one-dimensional sections do not remove the full-tensor degeneracy of the \(b=0\) potential; the uniaxial branch remains a selection assumption.

\section{Directional acoustic spectra and polarization tracking}
\label{sec:acoustics}
Figure~\ref{fig4} solves the full acoustic eigenproblem for the same three states and coefficients used in Fig.~\ref{fig3}, with \(\rho_0=1\). The radii in panels (a--c) are the actual phase speeds \(c_n/c_0\), with \(c_0=\sqrt{\mu/\rho_0}\), evaluated over the unit sphere using one common scale. Unlike the stiffness surfaces in Fig.~\ref{fig3}(d), these surfaces account for the eigenpolarizations. The isotropic transverse surfaces coincide everywhere, the uniaxial surfaces retain axial coincidence, and the biaxial surfaces resolve a more general angular pattern.

For the meridional cut \(\mathbf m=(\sin\vartheta,0,\cos\vartheta)\) and diagonal \(\mathbf Q\), the acoustic matrix separates into an out-of-plane sector and an in-plane block. The branches in Fig.~\ref{fig4}(d,e) are consequently identified by
\begin{equation}
	\rho_0c_{T,\rm out}^2=A_{22},\qquad
	\rho_0c_{L/T,\rm in}^2
	=\frac{A_{11}+A_{33}}2
	\mathbin{\pm}\sqrt{\left(\frac{A_{11}-A_{33}}2\right)^2+A_{13}^2}.
	\label{eq:figure_meridional_branches}
\end{equation}
The plus sign labels the quasi-longitudinal branch and the minus sign the in-plane quasi-transverse branch. Following these sectors preserves polarization identity when the two shear speeds cross; sorting solely by speed would exchange their labels. Panel (e) uses \(|\mathbf p_n\cdot\mathbf N|\), which is invariant under eigenvector sign reversal. In the biaxial case \(\mathbf N=\mathbf e_3\) is a reference axis, not a unique director. At exact degeneracy the individual transverse vectors depend on the chosen basis, whereas the projector onto their combined eigenspace is well defined.

The gap maps in Fig.~\ref{fig4}(f) distinguish protected zeros from state-dependent crossings. In the uniaxial case the gap vanishes at \(\mathbf m=\pm\mathbf N\) to numerical precision and reaches approximately \(0.33842c_0\) away from the axis. In the illustrated biaxial state, isolated crossings occur at \(\vartheta\simeq74.5559^\circ\) and \(105.4441^\circ\) in the meridional plane, together with their antipodal directions. A perturbation \(\delta\mathbf Q=\operatorname{diag}(0.015,-0.01,-0.005)\), which retains the \(D_2\) symmetry type, opens a gap of approximately \(0.00982c_0\) at the original crossing directions. This checks the lack of protection of those particular directions; it does not establish that all biaxial acoustic axes disappear. The full-sphere eigensystem residuals are of order \(10^{-15}\), and simultaneous rotation of the state and propagation direction obeys acoustic covariance to floating-point precision.

\section{Finite-amplitude gap checks and transition plots}
\label{sec:gaps}
Figure~\ref{fig5} tests these gap laws using complete acoustic eigenvalues rather than substituting the leading asymptotic expression for the numerical data. To distinguish the plotted observables, define
\begin{equation}
	\delta_T=\frac{|c_{T1}^2-c_{T2}^2|}{c_0^2},\qquad
	g_T=\frac{|c_{T1}-c_{T2}|}{c_0}
	=\frac{\Delta\omega_T}{kc_0},\qquad
	\delta_T=g_T\frac{c_{T1}+c_{T2}}{c_0}.
	\label{eq:figure_gap_observables}
\end{equation}
The quantity \(\delta_T\) is the difference of squared speeds, not the square of a speed difference. Panel (a), evaluated at \(s=0.01\), approaches \(\delta_T=|\gamma_2s|\sin^2\vartheta/(2\mu)+O(s^2)\); its largest discrepancy is approximately \(0.476\%\) of the first-order peak. Panels (b,c) recover a leading exponent near unity over \(10^{-5}\leq|s|\leq10^{-3}\), while the axial gap remains zero throughout the amplitude scan. Deviations at larger amplitudes reflect higher-order eigenvalue mixing, even though the stiffness itself is affine in \(\mathbf Q\).

At the equator, the propagation direction is a structural eigenvector and longitudinal--transverse mixing vanishes. For this particular potential the shear speeds are exactly
\begin{equation}
	c_{T,\rm in}^2=\frac{\mu+\gamma_2s/6}{\rho_0},\qquad
	c_{T,\rm out}^2=\frac{\mu-\gamma_2s/3}{\rho_0},\qquad
	\delta_T(\pi/2)=\frac{|\gamma_2s|}{2\mu},
	\label{eq:figure_equatorial_gap}
\end{equation}
within the stable regime in which these are the shear branches. This explains the exact linearity of panel (b) without extending it to every oblique direction. Panel (d) instead resolves the coupling selectivity: the normalized coefficient \(\delta_T/(|s|\sin^2\vartheta)\) approaches \(|\gamma_2|/(2\mu)\) as \(|s|\to0\), while its finite-amplitude dependence on \(\gamma_1\) comes from longitudinal--transverse mixing. In particular, setting \(\gamma_2=0\) removes the first-order gap but leaves the second-order prediction
\begin{equation}
	\delta_T=\frac{\gamma_1^2s^2}{\mu(\lambda+\mu)}
	\sin^2\vartheta\cos^2\vartheta+O(|s|^3),
	\qquad \gamma_2=0.
	\label{eq:figure_second_order_gap}
\end{equation}
This follows from the off-diagonal acoustic matrix element \(\gamma_1s\sin\vartheta\cos\vartheta\) and the isotropic longitudinal--transverse separation \(\lambda+\mu\). It demonstrates explicitly how an admissible symmetry reduction can produce \(p=2\), and why both the axis and equator retain a zero for this restricted coupling.

Figures~\ref{fig5}(e,f) insert the selected global equilibrium branch of Fig.~\ref{fig1}(d,e) into the same acoustic eigenproblem. The continuous case gives a fitted frequency-gap exponent of approximately \(0.50001\), consistent with Eq.~\eqref{eq:control_parameter_gap}; it tests the chosen mean-field uniaxial path rather than a universal transition exponent. In the first-order case, the equilibrium amplitude changes from zero to \(s_t=1/3\). At \(\vartheta=90^\circ\), the resulting jump is
\begin{equation}
	\frac{\Delta c_T^{\rm jump}}{c_0}
	=\left|\sqrt{1+\frac{\gamma_2s_t}{6\mu}}
	-\sqrt{1-\frac{\gamma_2s_t}{3\mu}}\right|
	\simeq0.119224,
	\label{eq:figure_first_order_spectral_jump}
\end{equation}
for the stated \(\gamma_2=1.4\) and \(\mu=1\). The two one-sided branches meet the coexistence point with different values; the jump is not located at a spinodal and is not produced by imposing a discontinuity on the acoustic model.

\section{Transient finite-element benchmark and synthetic inverse problem}
\label{sec:fem_inverse}
Figures~\ref{fig3}--\ref{fig5} establish the constitutive and spectral calculations, whereas Fig.~\ref{fig6}(a--e) supplies an independent transient displacement-field test. The FEM benchmark uses a homogeneous periodic strip \((x,z)\in[0,24]\times[0,0.96]\), retains all three displacement components, and assumes no variation along \(y\). Periodic transverse boundaries isolate bulk propagation, while damped end regions limit reflected arrivals. For each triangle, an engineering strain--displacement matrix \(\mathbf B_e\) and a stiffness Hessian evaluated independently from the scalar energy give
\begin{equation}
	\mathbf K_e=|\Omega_e|\mathbf B_e^T\mathbb C_V\mathbf B_e,
	\qquad
	\mathbf M\ddot{\mathbf u}+\mathbf D\dot{\mathbf u}+\mathbf K\mathbf u=\mathbf0.
	\label{eq:figure_fem_system}
\end{equation}
Here \(\mathbf M\) is the lumped mass matrix and \(\mathbf D\) acts only in the end sponges. The fine mesh has 19,200 linear triangles and 28,836 independent displacement degrees of freedom. Three-component Gaussian initial displacements are advanced by central differences without inserting acoustic eigenvectors or predicted arrival times into the initial field or solver. The geometry is a two-dimensional mesh with three-component motion, rather than a free-surface plate or a full three-dimensional finite-aperture beam model.

The waveforms and fields in Fig.~\ref{fig6}(b--d) correspond to propagation along \(\mathbf e_x\), with \(\mathbf N=(\cos\vartheta,0,\sin\vartheta)\) and \(s=0.9\). Thus \(\vartheta=0^\circ,45^\circ,90^\circ\) changes the angle between the wave and structural axis. The axial shear pulses coincide within discretization accuracy, whereas oblique propagation produces resolved shear arrivals. Peak times at receivers \(x_1=7\) and \(x_2=13\) yield
\begin{equation}
	c_n^{\rm TOF}=\frac{x_2-x_1}{t_{n,2}^{\rm peak}-t_{n,1}^{\rm peak}},
	\qquad
	\varepsilon_{c,n}=\frac{|c_n^{\rm TOF}-c_n^{\rm ac}|}{c_n^{\rm ac}}.
	\label{eq:figure_tof_speed}
\end{equation}
The peak estimator measures pulse transit speed. For a differentiable acoustic branch, $\omega_n=k c_n(\mathbf m)$ implies
\begin{equation}
	\mathbf v_{g,n}=\nabla_{\mathbf k}\omega_n
	=c_n\mathbf m+(\mathbf I-\mathbf m\otimes\mathbf m)\nabla_{\mathbf m}c_n,
	\qquad \mathbf m\cdot\mathbf v_{g,n}=c_n.
	\label{eq:group_velocity}
\end{equation}
Thus nondispersive frequency dependence does not imply equality of the phase- and group-velocity vectors. The transversely uniform periodic-strip benchmark compares axial transit with the phase speed for the same wavevector; it does not model the ray direction of a finite-aperture beam. Comparison with acoustic phase speed is appropriate here because the physical homogeneous frozen-state model is nondispersive; the residual also includes mesh and time discretization effects. Across the nine angle--branch combinations in Fig.~\ref{fig6}(e), the maximum relative error is \(0.2133\%\). The measured shear-speed differences at \(45^\circ\) and \(90^\circ\) are approximately \(0.04714c_0\) and \(0.33784c_0\), compared with acoustic predictions of \(0.04687c_0\) and \(0.33842c_0\). Refinement from \(h=0.12\) through \(0.06\) to \(0.03\) reduces the maximum error in the equatorial case from \(1.3389\%\) to \(0.6760\%\) and \(0.2133\%\), providing a discretization check independent of the symmetry argument.

Figure~\ref{fig6}(f) examines a separate synthetic inverse problem. The true state is the \(45^\circ\) tensor, and all five coefficients in the orthonormal expansion \(\mathbf Q=\sum_Aq_A\mathbf T_A\) are fitted without imposing uniaxiality. Twenty-four directions distributed over a hemisphere supply three acoustic speeds each, with the elastic coefficients and density calibrated and held fixed. Independent relative Gaussian speed noise is added before sorting each measured triplet. With \(\Lambda_{mn}(\mathbf q)\) denoting an acoustic eigenvalue at direction \(m\), the fitted objective and relative speed sensitivity are
\begin{align}
	\Phi(\mathbf q)&=\frac12\sum_{m,n}r_{mn}(\mathbf q)^2,\qquad
	r_{mn}=\frac{\Lambda_{mn}(\mathbf q)-\rho_0(c_{mn}^{\rm meas})^2}
	{2\rho_0(c_{mn}^{\rm meas})^2},
	\label{eq:figure_inverse_objective}\\
	\mathcal J^{\rm rel}_{mn,A}
	&=\frac{1}{c_{mn}}\frac{\partial c_{mn}}{\partial q_A}
	=\frac{\mathbf p_{mn}^T(\partial\mathsf A_m/\partial q_A)\mathbf p_{mn}}
	{2\Lambda_{mn}},\qquad
	\kappa(\mathcal J^{\rm rel})=\frac{\sigma_{\max}}{\sigma_{\min}}.
	\label{eq:figure_inverse_sensitivity}
\end{align}
The selected directions avoid exact acoustic degeneracy at the true state, so the simple-eigenvalue sensitivity is well defined. The 24-direction matrix has rank five and condition number approximately \(1.411\). One direction has rank three; the chosen two generic directions already give full local rank, but this does not imply global uniqueness from every two-direction design. The favorable conditioning refers to this observation normalization, orthonormal parameter basis, and calibrated model.

For 100 independent noise realizations at each level, with four non-truth initial guesses per fit, the median relative Frobenius reconstruction errors are \(2.041\%\), \(4.275\%\), and \(12.052\%\) for speed noise of \(1\%\), \(2\%\), and \(5\%\), respectively. The corresponding 95th-percentile errors are \(3.033\%\), \(6.432\%\), and \(18.475\%\). Slight biaxiality in a noisy estimate is permitted by the five-component fit; it is not evidence of a true symmetry transition. These synthetic observations are generated by the calibrated acoustic model, independently of the three FEM loading cases. Together the two tests demonstrate forward agreement and conditional inverse recovery for this benchmark. Experimental validation remains to combine independently measured structural orientation statistics with polarization-resolved wave speeds under controlled loading or temperature.

\section{Frozen and relaxed incremental response}
\label{sec:dynamic}
The difference between frozen and relaxed response can be displayed without changing the structural theory. Linearize a stable homogeneous equilibrium at fixed temperature, represent structural increments in the five-component basis above, and write the quadratic energy as
\begin{equation}
	\delta^2\Psi=\frac12\delta\mathbf E:\mathbb C_f:\delta\mathbf E
	+\delta\mathbf q\cdot\mathbb B_{qE}:\delta\mathbf E
	+\frac12\delta\mathbf q\cdot\mathbb H_q\delta\mathbf q.
	\label{eq:coupled_incremental_energy}
\end{equation}
For positive-definite mobility \(\mathbb M_q\), a harmonic increment with convention \(e^{-i\omega t}\) obeys \((\mathbb H_q-i\omega\mathbb M_q^{-1})\delta\mathbf q=-\mathbb B_{qE}:\delta\mathbf E\). Eliminating the structural increment gives
\begin{equation}
	\mathbb C_{\rm eff}(\omega)=\mathbb C_f
	-\mathbb B_{qE}^{T}(\mathbb H_q-i\omega\mathbb M_q^{-1})^{-1}\mathbb B_{qE}.
	\label{eq:relaxed_dynamic_tangent}
\end{equation}
For a strictly stable structural subspace, $\mathbb H_q\succ0$, and a positive-definite mobility, this elimination is compatible with dissipation. With the $e^{-i\omega t}$ convention, $-\operatorname{Im}\mathbb C_{\rm eff}(\omega)$ is positive semidefinite for $\omega>0$, giving nonnegative cycle-averaged dissipation. A stable fully relaxed equilibrium also requires the static Schur complement $\mathbb C_f-\mathbb B_{qE}^T\mathbb H_q^{-1}\mathbb B_{qE}$ to be positive on admissible strain increments. Structural stability alone does not establish this additional mechanical condition.

At high frequency this approaches the frozen tangent; at low frequency it gives the relaxed Schur-complement tangent when the structural Hessian is invertible. At nonzero spatial wavenumber, the isotropic gradient penalty adds \(\kappa_Q|\mathbf q|^2\mathbb I\) to the structural Hessian. Zero-curvature orientation modes require treating the appropriate constrained subspace rather than inserting a singular static inverse. For the particular energy in Eq.~\eqref{eq:minimal_energy} at zero base strain, \(\mathbb B_{qE}=\mathbf0\), so this linear relaxation correction vanishes: the probe does not drive a first-order structural change. It can be present under prestrain, transformation-strain coupling, or additional mixed terms. The conservative acoustic eigenproblem above is therefore justified for its stated example, while the more general dynamic problem can contain attenuation and frequency-dependent moduli.

\section{Orbit geometry and invariant classification}
\label{app:orbit_geometry}

The orbit of \(\mathbf Q\) is \(\mathcal O_{\mathbf Q}=\{g^T\mathbf Qg:g\in G_0\}\). In regular cases,
\begin{equation}
	\mathcal O_{\mathbf Q}\simeq G_0/H_{\mathbf Q},
	\qquad
	\dim\mathcal O_{\mathbf Q}=\dim G_0-\dim H_{\mathbf Q}.
	\label{eq:orbit_dimension}
\end{equation}
For \(G_0=SO(3)\), the isotropic, uniaxial, and biaxial orbit dimensions are \(0\), \(2\), and \(3\), respectively. Orbit dimension counts orientation degrees of freedom generated by the parent group, not all thermodynamic degrees of freedom.

For a traceless \(3\times3\) symmetric tensor, the characteristic polynomial is
\begin{equation}
	\chi_{\mathbf Q}(q)=q^3-\frac12J_2q-\frac13J_3.
	\label{eq:Q_characteristic}
\end{equation}
Reality of its eigenvalues implies the discriminant condition
\begin{equation}
	J_2\geq0,
	\qquad 6J_3^2\leq J_2^3.
	\label{eq:orbit_domain}
\end{equation}
Equality holds for uniaxial states, the origin represents the isotropic state, and strict inequality describes generic biaxial states. Equation~\eqref{eq:biaxiality} is a normalized measure of biaxiality relative to this boundary in invariant coordinates; it is not a metric distance.

\section{Dissipation inequality for the structural field}
\label{app:thermodynamics}

For an arbitrary material part \(\mathcal P_0\), include the structural microtraction \(\mathbf t_Q\) in the free-energy imbalance. The terms involving \(\mathbf Q\) are
\begin{equation}
	\mathfrak D_Q=-\int_{\mathcal P_0}\left[
	\frac{\partial\Psi}{\partial\mathbf Q}:\dot{\mathbf Q}
	+\mathbf M_Q\mathbin{\vdots}\operatorname{Grad}\dot{\mathbf Q}\right]\mathrm dV
	+\int_{\partial\mathcal P_0}\mathbf t_Q:\dot{\mathbf Q}\,\mathrm dA.
	\label{eq:global_Q_dissipation}
\end{equation}
Integration by parts gives
\begin{equation}
	-\int_{\mathcal P_0}\mathbf M_Q\mathbin{\vdots}\operatorname{Grad}\dot{\mathbf Q}\,\mathrm dV
	=\int_{\mathcal P_0}\operatorname{Div}\mathbf M_Q:\dot{\mathbf Q}\,\mathrm dV
	-\int_{\partial\mathcal P_0}(\mathbf M_Q\mathbf N):\dot{\mathbf Q}\,\mathrm dA.
	\label{eq:integration_by_parts}
\end{equation}
Choosing the energetic microtraction \(\mathbf t_Q=\mathbf M_Q\mathbf N\) leaves the local inequality
\begin{equation}
	\mathfrak d_Q=\left(-\frac{\partial\Psi}{\partial\mathbf Q}
	+\operatorname{Div}\mathbf M_Q\right):\dot{\mathbf Q}\geq0.
	\label{eq:local_Q_dissipation}
\end{equation}
Since \(\dot{\mathbf Q}\in\operatorname{Sym}_0(3)\), replacing the force by its projection does not change the power. Equation~\eqref{eq:kinetic_law} then satisfies the inequality by construction. The natural boundary condition is \(\mathbf M_Q\mathbf N=\mathbf0\); a Dirichlet condition may instead prescribe \(\mathbf Q\).

\section{First-order splitting of the transverse eigenspace}
\label{app:perturbation}

Let \(P_T=\mathbf I-\widehat{\mathbf k}\otimes\widehat{\mathbf k}\) project onto the isotropic transverse eigenspace. The first-order corrections to the repeated eigenvalue \(\mu\) are the eigenvalues of
\begin{equation}
	K^{(T)}=P_T\mathsf A_1P_T\big|_{\operatorname{Range}P_T}.
	\label{eq:projected_perturbation_appendix}
\end{equation}
In any orthonormal transverse basis, these are
\begin{equation}
	\eta_\pm=\frac{K_{11}^{(T)}+K_{22}^{(T)}}{2}
	\pm\sqrt{\left(\frac{K_{11}^{(T)}-K_{22}^{(T)}}{2}\right)^2+(K_{12}^{(T)})^2}.
	\label{eq:eta_pm}
\end{equation}
Their difference is basis independent. Expanding \(c_{T\pm}=\sqrt{(\mu+s\eta_\pm)/\rho_0}\) gives Eq.~\eqref{eq:frequency_split}. If the little group acts irreducibly on the transverse plane, a symmetric operator commuting with the group must be proportional to the identity there; consequently \(\eta_+=\eta_-\), and the linear splitting vanishes.

\printcredits
\bibliographystyle{cas-model2-names}
\bibliography{references}

@article{Zheng1994,
	author  = {Zheng, Q.-S. and Boehler, J. P.},
	title   = {The description, classification, and reality of material and physical symmetries},
	journal = {Acta Mech.},
	year    = {1994},
	volume  = {105},
	pages   = {73--89}
}

@article{Gross1996,
	author  = {Gross, D. J.},
	title   = {The role of symmetry in fundamental physics},
	journal = {Proc. Natl. Acad. Sci. U.S.A.},
	year    = {1996},
	volume  = {93},
	pages   = {14256--14259}
}

@article{Trovalusci1999,
	author  = {Trovalusci, P. and Masiani, R.},
	title   = {Material symmetries of micropolar continua equivalent to lattices},
	journal = {Int. J. Solids Struct.},
	year    = {1999},
	volume  = {36},
	pages   = {2091--2108}
}

@article{Brovko2021,
	author  = {Brovko, G. L.},
	title   = {Objective Tensors and Their Mappings in Classical Continuum Mechanics},
	journal = {Mech. Solids},
	year    = {2021},
	volume  = {56},
	pages   = {65--83}
}

@article{Davini2001,
	author  = {Davini, C.},
	title   = {Some remarks on the continuum theory of defects in solids},
	journal = {Int. J. Solids Struct.},
	year    = {2001},
	volume  = {38},
	pages   = {1169--1182}
}

@article{Carroll2001,
	author  = {Carroll, M. M.},
	title   = {A representation theorem for volume-preserving transformations},
	journal = {Int. J. Non-Linear Mech.},
	year    = {2001},
	volume  = {39},
	pages   = {219--224}
}

@article{Shai2001,
	author  = {Shai, O.},
	title   = {Deriving structural theorems and methods using Tellegen's theorem and combinatorial representations},
	journal = {Int. J. Solids Struct.},
	year    = {2001},
	volume  = {38},
	pages   = {8037--8052}
}

@article{Rice1971,
	author  = {Rice, J. R.},
	title   = {Inelastic constitutive relations for solids: An internal-variable theory and its application to metal plasticity},
	journal = {J. Mech. Phys. Solids},
	year    = {1971},
	volume  = {19},
	pages   = {433--455}
}

@article{Resende1987,
	author  = {Resende, L.},
	title   = {A damage mechanics constitutive theory for the inelastic behaviour of concrete},
	journal = {Comput. Methods Appl. Mech. Eng.},
	year    = {1987},
	volume  = {60},
	pages   = {57--93}
}

@article{Segev1986,
	author  = {Segev, R.},
	title   = {Forces and the existence of stresses in invariant continuum mechanics},
	journal = {J. Math. Phys.},
	year    = {1986},
	volume  = {27},
	pages   = {163--170}
}

@article{Daymond2007,
	author  = {Daymond, M. R. and Young, M. L. and Almer, J. D. and Dunand, D. C.},
	title   = {Strain and texture evolution during mechanical loading of a crack tip in martensitic shape-memory {NiTi}},
	journal = {Acta Mater.},
	year    = {2007},
	volume  = {55},
	pages   = {3929--3942}
}

@article{Yi2006,
	author  = {Yi, S.-B. and Davies, C. H. J. and Brokmeier, H.-G. and Bolmaro, R. E. and Kainer, K. U. and Homeyer, J.},
	title   = {Deformation and texture evolution in {AZ31} magnesium alloy during uniaxial loading},
	journal = {Acta Mater.},
	year    = {2006},
	volume  = {54},
	pages   = {549--562}
}

@article{Stupkiewicz2002,
	author  = {Stupkiewicz, S. and Petryk, H.},
	title   = {Modelling of laminated microstructures in stress-induced martensitic transformations},
	journal = {J. Mech. Phys. Solids},
	year    = {2002},
	volume  = {50},
	pages   = {2303--2331}
}

@article{Lu1997,
	author  = {Lu, Z. K. and Weng, G. J.},
	title   = {Martensitic transformation and stress-strain relations of shape-memory alloys},
	journal = {J. Mech. Phys. Solids},
	year    = {1997},
	volume  = {45},
	pages   = {1905--1921}
}

@article{Litewka2003,
	author  = {Litewka, A. and Debinski, J.},
	title   = {Load-induced oriented damage and anisotropy of rock-like materials},
	journal = {Int. J. Plast.},
	year    = {2003},
	volume  = {19},
	pages   = {2171--2191}
}

@article{Margueres2013,
	author  = {Marguères, P. and Meraghni, F.},
	title   = {Damage induced anisotropy and stiffness reduction evaluation in composite materials using ultrasonic wave transmission},
	journal = {Compos. Part A Appl. Sci. Manuf.},
	year    = {2013},
	volume  = {45},
	pages   = {134--144}
}

@article{Ambrosi2011,
	author  = {Ambrosi, D. and Ateshian, G. A. and Arruda, E. M. and Cowin, S. C. and Dumais, J. and Goriely, A. and Holzapfel, G. A. and Humphrey, J. D. and Kemkemer, R. and Kuhl, E. and Olberding, J. E. and Taber, L. A. and Garikipati, K.},
	title   = {Perspectives on biological growth and remodeling},
	journal = {J. Mech. Phys. Solids},
	year    = {2011},
	volume  = {59},
	pages   = {863--883}
}

@article{Berlincourt1959,
	author  = {Berlincourt, D. and Krueger, H. H. A.},
	title   = {Domain processes in lead titanate zirconate and barium titanate ceramics},
	journal = {J. Appl. Phys.},
	year    = {1959},
	volume  = {30},
	pages   = {1804--1810}
}

@article{Kreher2002,
	author  = {Kreher, W. S.},
	title   = {Influence of domain switching zones on the fracture toughness of ferroelectrics},
	journal = {J. Mech. Phys. Solids},
	year    = {2002},
	volume  = {50},
	pages   = {1029--1050}
}

@article{Li2009,
	author  = {Li, X. and Yu, H. S. and Li, X. S.},
	title   = {Macro--micro relations in granular mechanics},
	journal = {Int. J. Solids Struct.},
	year    = {2009},
	volume  = {46},
	pages   = {4331--4341}
}

@article{Li2015,
	author  = {Li, X. S. and Dafalias, Y. F.},
	title   = {Dissipation consistent fabric tensor definition from {DEM} to continuum for granular media},
	journal = {J. Mech. Phys. Solids},
	year    = {2015},
	volume  = {78},
	pages   = {141--153}
}

@article{Jack2007,
	author  = {Jack, D. A. and Smith, D. E.},
	title   = {The effect of fibre orientation closure approximations on mechanical property predictions},
	journal = {Compos. Part A Appl. Sci. Manuf.},
	year    = {2007},
	volume  = {38},
	pages   = {975--982}
}

@article{Rosakis2000,
	author  = {Rosakis, P. and Rosakis, A. J. and Ravichandran, G. and Hodowany, J.},
	title   = {A thermodynamic internal variable model for the partition of plastic work into heat and stored energy in metals},
	journal = {J. Mech. Phys. Solids},
	year    = {2000},
	volume  = {48},
	pages   = {581--607}
}

@article{Svolos2025,
	author  = {Svolos, L. and Tran, Q.-T. and Boureima, I. D. and Anghel, V. and Garikipati, K. and Mourad, H. M.},
	title   = {A phase-field fracture formulation for generalized standard materials: The interplay between thermomechanics and damage},
	journal = {J. Mech. Phys. Solids},
	year    = {2025},
	volume  = {201},
	pages   = {106154}
}

@article{Wu2017,
	author  = {Wu, J.-Y.},
	title   = {A unified phase-field theory for the mechanics of damage and quasi-brittle failure},
	journal = {J. Mech. Phys. Solids},
	year    = {2017},
	volume  = {103},
	pages   = {72--99}
}

@book{Landau1982,
	author    = {Landau, L. D. and Lifshitz, E. M.},
	title     = {Mechanics},
	publisher = {Elsevier},
	year      = {1982},
	volume    = {1}
}

@article{Arsenlis2002,
	author  = {Arsenlis, A. and Parks, D. M.},
	title   = {Modeling the evolution of crystallographic dislocation density in crystal plasticity},
	journal = {J. Mech. Phys. Solids},
	year    = {2002},
	volume  = {50},
	pages   = {1979--2009}
}

@article{Acharya2001,
	author  = {Acharya, A.},
	title   = {A model of crystal plasticity based on the theory of continuously distributed dislocations},
	journal = {J. Mech. Phys. Solids},
	year    = {2001},
	volume  = {49},
	pages   = {761--784}
}

@article{Tommasi2009,
	author  = {Tommasi, A. and Knoll, M. and Vauchez, A. and Signorelli, J. W. and Thoraval, C. and Logé, R.},
	title   = {Structural reactivation in plate tectonics controlled by olivine crystal anisotropy},
	journal = {Nat. Geosci.},
	year    = {2009},
	volume  = {2},
	pages   = {423--427}
}

@article{Kobayashi2025,
	author  = {Kobayashi, T. and Nishioka, Y. and Otsuka, H. and Tanimoto, M.},
	title   = {More about quark {Yukawa} textures from selection rules without group actions},
	journal = {J. High Energy Phys.},
	year    = {2025},
	volume  = {2025},
	pages   = {177}
}

@article{Lami2024,
	author  = {Lami, G. and Collura, M.},
	title   = {Unveiling the stabilizer group of a matrix product state},
	journal = {Phys. Rev. Lett.},
	year    = {2024},
	volume  = {133},
	pages   = {010602}
}

@article{Liu2026,
	author  = {Liu, Y. and Liu, C.-C.},
	title   = {Antiferroaxial altermagnetism},
	journal = {Phys. Rev. Lett.},
	year    = {2026},
	volume  = {136},
	pages   = {256709}
}

@article{Dao2023,
	author  = {Dao, C. and Everts, J. C. and Ravnik, M. and Tserkovnyak, Y.},
	title   = {Nematronics: Reciprocal coupling between ionic currents and nematic dynamics},
	journal = {Phys. Rev. Lett.},
	year    = {2023},
	volume  = {130},
	pages   = {168102}
}

@article{Coelho2023,
	author  = {Coelho, R. C. V. and Zhao, H. and Tasinkevych, M. and Smalyukh, I. I. and Telo da Gama, M. M.},
	title   = {Sculpting liquid crystal skyrmions with external flows},
	journal = {Phys. Rev. Res.},
	year    = {2023},
	volume  = {5},
	pages   = {033210}
}

@article{Liu2016,
	author  = {Liu, K. and Nissinen, J. and Slager, R.-J. and Wu, K. and Zaanen, J.},
	title   = {Generalized liquid crystals: Giant fluctuations and the vestigial chiral order of {$I$}, {$O$}, and {$T$} matter},
	journal = {Phys. Rev. X},
	year    = {2016},
	volume  = {6},
	pages   = {041025}
}

@article{Nissinen2016,
	author  = {Nissinen, J. and Liu, K. and Slager, R.-J. and Wu, K. and Zaanen, J.},
	title   = {Classification of point-group-symmetric orientational ordering tensors},
	journal = {Phys. Rev. E},
	year    = {2016},
	volume  = {94},
	pages   = {022701}
}

@article{Menzel2003,
	author  = {Menzel, A. and Steinmann, P.},
	title   = {A view on anisotropic finite hyper-elasticity},
	journal = {Eur. J. Mech. A Solids},
	year    = {2003},
	volume  = {22},
	pages   = {71--87}
}

@article{Karateev2019,
	author  = {Karateev, D. and Kravchuk, P. and Simmons-Duffin, D.},
	title   = {Harmonic analysis and mean field theory},
	journal = {J. High Energy Phys.},
	year    = {2019},
	volume  = {2019},
	pages   = {217}
}

@article{Benallal2004,
	author  = {Benallal, A. and Bigoni, D.},
	title   = {Effects of temperature and thermo-mechanical couplings on material instabilities and strain localization of inelastic materials},
	journal = {J. Mech. Phys. Solids},
	year    = {2004},
	volume  = {52},
	pages   = {725--753}
}

@article{Moulton2011,
	author  = {Moulton, D. E. and Goriely, A.},
	title   = {Circumferential buckling instability of a growing cylindrical tube},
	journal = {J. Mech. Phys. Solids},
	year    = {2011},
	volume  = {59},
	pages   = {525--537}
}

@article{Wang2024,
	author  = {Wang, Z.-Y. and Wu, Z.-J. and Duan, G. and Shui, L.-Q. and Li, Y.-W.},
	title   = {Extracting longitudinal waves by utilizing the symmetry of piezoelectric sphere},
	journal = {Int. J. Mech. Sci.},
	year    = {2024},
	volume  = {279},
	pages   = {109499}
}

@article{Liang2025,
	author  = {Liang, S.-N. and Xie, J.-L. and He, C. and Yu, S.-Y. and Chen, Y.-F.},
	title   = {Manipulating Mobius edge states in elastic wave phononic crystals},
	journal = {Phys. Rev. B},
	year    = {2025},
	volume  = {111},
	pages   = {184103}
}

@article{Liu2018,
	author  = {Liu, T.-W. and Semperlotti, F.},
	title   = {Tunable acoustic Valley--Hall edge states in reconfigurable phononic elastic waveguides},
	journal = {Phys. Rev. Appl.},
	year    = {2018},
	volume  = {9},
	pages   = {014001}
}

@article{Goryachev2021,
	author  = {Goryachev, M. and Campbell, W. M. and Heng, I. S. and Galliou, S. and Ivanov, E. N. and Tobar, M. E.},
	title   = {Rare events detected with a bulk acoustic wave high frequency gravitational wave antenna},
	journal = {Phys. Rev. Lett.},
	year    = {2021},
	volume  = {127},
	pages   = {071102}
}

@article{Csaki2021,
	author  = {Csáki, C. and Hong, S. and Shirman, Y. and Telem, O. and Terning, J.},
	title   = {Completing multiparticle representations of the {Poincaré} group},
	journal = {Phys. Rev. Lett.},
	year    = {2021},
	volume  = {127},
	pages   = {041601}
}

@article{Zhang2014,
	author  = {Zhang, F. and Kane, C. L.},
	title   = {Time-reversal-invariant {$Z_4$} fractional {Josephson} effect},
	journal = {Phys. Rev. Lett.},
	year    = {2014},
	volume  = {113},
	pages   = {036401}
}

@article{Weissenhofer2025,
	author  = {Weißenhofer, M. and Rieger, P. and Mrudul, M. S. and Mikadze, L. and Nowak, U. and Oppeneer, P. M.},
	title   = {Chiral phonons arising from chirality-selective magnon-phonon coupling},
	journal = {Phys. Rev. Lett.},
	year    = {2025},
	volume  = {135},
	pages   = {216701}
}

@article{Giljean2024,
	author  = {Giljean, S. and Gaillard, Y. and Rousselot, C. and Bouillet, C. and Lenertz, M. and Richard, F. and Pac, M.-J.},
	title   = {A mechanical {TiAl/TiAlN} multinanolayer coating model based on microstructural analysis and nanoindentation},
	journal = {Acta Mater.},
	year    = {2024},
	volume  = {280},
	pages   = {120312}
}

@article{Lan2018,
	author  = {Lan, B. and Britton, T. B. and Jun, T.-S. and Gan, W. and Hofmann, M. and Dunne, F. P. E. and Lowe, M. J. S.},
	title   = {Direct volumetric measurement of crystallographic texture using acoustic waves},
	journal = {Acta Mater.},
	year    = {2018},
	volume  = {159},
	pages   = {384--394}
}

@article{Castellano2017,
	author  = {Castellano, A. and Fraddosio, A. and Piccioni, M. D.},
	title   = {Ultrasonic goniometric immersion tests for the characterization of fatigue post-{LVI} damage induced anisotropy superimposed to the constitutive anisotropy of polymer composites},
	journal = {Compos. Part B Eng.},
	year    = {2017},
	volume  = {116},
	pages   = {122--136}
}

@article{Zhu2024,
	author  = {Zhu, P. and Zhang, H. and Santulli, C. and Sfarra, S. and Usamentiaga, R. and Vavilov, V.P. and Maldague, X.},
	title   = {Contactless and nondestructive evaluation of residual stress distribution in modified and pure {HDPE} materials using a novel terahertz method and line-scan thermographic technique},
	journal = {Compos. Part A Appl. Sci. Manuf.},
	year    = {2024},
	volume  = {183},
	pages   = {108220}
}

@article{Zhu2026a,
    author  = {Zhu, P. and Lecompagnon, J. and Hirsch, P.D. and Ziegler, M.},
    title   = {Spectral Geometry and Dispersion-Constrained Projection of Diffusive Fields},
    journal = {arXiv preprint},
    year    = {2026},
    eprint  = {2609.09916},
    archivePrefix = {arXiv},
    primaryClass  = {physics.app-ph}
}

@article{Zhu2026b,
    author  = {Zhu, P.},
    title   = {Discovering Hidden Predictive States in Kinetic Transport},
    journal = {arXiv preprint},
    year    = {2026},
    eprint  = {2610.28383},
    archivePrefix = {arXiv},
    primaryClass  = {physics.app-ph}
}

@article{Zhu2024_2,
	author  = {Zhu, P. and Wang, R. and Sfarra, S. and Vavilov, V. and Maldague, X. and Zhang, H.},
	title   = {A novel force-temperature model for evaluating internal forces in {CFRP} by means of infrared thermography},
	journal = {NDT E Int.},
	year    = {2024},
	volume  = {143},
	pages   = {103066}
}

@article{Zhu2026,
	author  = {Zhu, P. and Zhang, H. and Sfarra, S. and Sarasini, F. and Usamentiaga, R. and Steenackers, G. and Ibarra-Castanedo, C. and Maldague, X.},
	title   = {Thermal diffusivity characterization of impacted composites using evaporative cryocooling excitation and inverse physics-informed neural networks},
	journal = {IEEE Trans. Instrum. Meas.},
	year    = {2026},
	volume  = {75},
	pages   = {6003011}
}

@article{Zhu2025,
	author  = {Zhu, P. and Wang, R. and Sivagurunathan, K. and Sfarra, S. and Sarasini, F. and Ibarra-Castanedo, C. and Maldague, X. and Zhang, H. and Mandelis, A.},
	title   = {Frequency multiplexed photothermal correlation tomography for non-destructive evaluation of manufactured materials},
	journal = {Int. J. Extrem. Manuf.},
	year    = {2025},
	volume  = {7},
	pages   = {035601}
}

@article{Zhu2025_2,
	author  = {Zhu, P. and Zhang, H. and Sfarra, S. and Sarasini, F. and Usamentiaga, R. and Steenackers, G. and Ibarra-Castanedo, C. and Maldague, X.},
	title   = {A comprehensive evaluation of the low-velocity impact behaviour of intraply hybrid flax/basalt composites using infrared thermography and terahertz time-domain spectroscopy techniques},
	journal = {NDT E Int.},
	year    = {2025},
	volume  = {154},
	pages   = {103361}
}

@article{Zhu2026_2,
	author  = {Zhu, P. and Zhang, H. and Sfarra, S. and Sarasini, F. and Maldague, X. and Mandelis, A.},
	title   = {Three-dimensional wide-bandwidth quantum energy truncation terahertz coherence tomography},
	journal = {Phys. Rev. Lett.},
	year    = {2026},
	volume  = {137},
	pages   = {026901}
}

@article{Cowin1987,
 author={Cowin, Stephen C. and Mehrabadi, Morteza M.},
 title={On the identification of material symmetry for anisotropic elastic materials},
 journal={The Quarterly Journal of Mechanics and Applied Mathematics},
 year={1987},volume={40},number={4},pages={451--476},doi={10.1093/qjmam/40.4.451}
}
\end{document}